\documentclass[12pt,preprint]{aastex}

\slugcomment{Submitted to ...}

\shorttitle{Legacy ExtraGalactic UV Survey (LEGUS)}
\shortauthors{Calzetti et al.}

\begin{document}

\title{Legacy ExtraGalactic UV Survey (LEGUS) with The Hubble Space Telescope. 
I. Survey Description.\altaffilmark{1}}

\author{D. Calzetti\altaffilmark{2}, J.C. Lee\altaffilmark{3, 4}, E. Sabbi\altaffilmark{3}, A. Adamo\altaffilmark{5}, L.J. Smith\altaffilmark{6}, J.E. Andrews\altaffilmark{2, 7}, L. Ubeda\altaffilmark{3}, S.N. Bright\altaffilmark{3}, D. Thilker\altaffilmark{8}, A. Aloisi\altaffilmark{3}, T.M. Brown\altaffilmark{3}, R. Chandar\altaffilmark{9}, 
C. Christian \altaffilmark{3}, M. Cignoni\altaffilmark{3}, G.C. Clayton\altaffilmark{10}, R. da Silva\altaffilmark{11},  S.E. de Mink\altaffilmark{12,13,14,*}, C. Dobbs\altaffilmark{15}, B.G. Elmegreen\altaffilmark{16}, D.M. Elmegreen\altaffilmark{17}, A.S. Evans\altaffilmark{18,19}, M. Fumagalli\altaffilmark{20,13}, J.S. Gallagher III\altaffilmark{21}, D.A. Gouliermis\altaffilmark{22,23}, E.K. Grebel\altaffilmark{24}, A. Herrero\altaffilmark{25,26}, D.A. Hunter\altaffilmark{27}, K.E. Johnson\altaffilmark{18}, R.C. Kennicutt \altaffilmark{28}, H. Kim\altaffilmark{29,30},  M.R. Krumholz\altaffilmark{11}, D. Lennon\altaffilmark{31}, K. Levay\altaffilmark{3}, C. Martin\altaffilmark{32}, P. Nair\altaffilmark{33}, A. Nota\altaffilmark{6}, G. \"Ostlin\altaffilmark{5}, A. Pellerin\altaffilmark{34}, J. Prieto\altaffilmark{35}, M.W. Regan\altaffilmark{3}, J.E. Ryon\altaffilmark{21},  D. Schaerer\altaffilmark{36}, D. Schiminovich\altaffilmark{37}, M. Tosi\altaffilmark{38}, S.D. Van Dyk\altaffilmark{39},  R. Walterbos\altaffilmark{40}, B.C. Whitmore\altaffilmark{3}, A. Wofford\altaffilmark{41}}

\altaffiltext{1}{Based on observations obtained with the NASA/ESA Hubble Space Telescope, at the Space Telescope Science Institute, which is operated by the Association of Universities for Research in Astronomy, Inc., under NASA contract NAS 5-26555. } 
\altaffiltext{2}{Dept. of Astronomy, University of Massachusetts -- Amherst, Amherst, MA 01003;\\ calzetti@astro.umass.edu}
\altaffiltext{3}{Space Telescope Science Institute, Baltimore, MD}
\altaffiltext{4}{Visiting Astronomer, Spitzer Science Center, Caltech. Pasadena, CA}
\altaffiltext{5}{Dept. of Astronomy, The Oskar Klein Centre, Stockholm University, Stockholm, Sweden }
\altaffiltext{6}{European Space Agency/Space Telescope Science Institute, Baltimore, MD}
\altaffiltext{7}{Dept. of Astronomy, University of Arizona, Tucson, AZ}
\altaffiltext{8}{Dept. of Physics and Astronomy, The Johns Hopkins University, Baltimore, MD}
\altaffiltext{9}{Dept. of Physics and Astronomy, University of Toledo, Toledo, OH}
\altaffiltext{10}{Dept. of Physics and Astronomy, Louisiana State University, Baton Rouge, LA}
\altaffiltext{11}{Dept. of Astronomy \& Astrophysics, University of California -- Santa Cruz, Santa Cruz, CA}
\altaffiltext{12}{Astronomical Institute Anton Pannekoek, Amsterdam University, Amsterdam, The Netherlands}
\altaffiltext{13}{Observatories of the Carnegie Institution for Science, Pasadena, CA}
\altaffiltext{14}{TAPIR institute, California Institute of Technology, Pasadena, CA}
\altaffiltext{15}{School of Physics and Astronomy, University of Exeter, Exeter, United Kingdom}
\altaffiltext{16}{IBM Research Division, T.J. Watson Research Center, Yorktown Hts., NY}
\altaffiltext{17}{Dept. of Physics and Astronomy, Vassar College, Poughkeepsie, NY}
\altaffiltext{18}{Dept. of Astronomy, University of Virginia, Charlottesville, VA}
\altaffiltext{19}{National Radio Astronomy Observatory, Charlottesville, VA}
\altaffiltext{20}{Institute for Computational Cosmology, Department of Physics, Durham University, Durham, United Kingdom}
\altaffiltext{21}{Dept. of Astronomy, University of Wisconsin--Madison, Madison, WI}
\altaffiltext{22}{Centre for Astronomy, Institute for Theoretical Astrophysics, University of Heidelberg, Heidelberg, Germany}
\altaffiltext{23}{Max Planck Institute for Astronomy, Heidelberg, Germany}
\altaffiltext{24}{Astronomisches Rechen-Institut, Zentrum f\"ur Astronomie der Universit\"at Heidelberg, Heidelberg, Germany}
\altaffiltext{25}{Instituto de Astrofisica de Canarias, La Laguna, Tenerife, Spain}
\altaffiltext{26}{Departamento de Astrofisica, Universidad de La Laguna, Tenerife, Spain}
\altaffiltext{27}{Lowell Observatory, Flagstaff, AZ}
\altaffiltext{28}{Institute of Astronomy, University of Cambridge, Cambridge, United Kingdom}
\altaffiltext{29}{School of Earth and Space Exploration, Arizona State University, Tempe, AZ}
\altaffiltext{30}{Korea Astronomy and Space Science Institute, Daejeon, Republic of Korea}
\altaffiltext{31}{European Space Astronomy Centre, ESA, Villanueva de la Ca\~nada, Madrid, Spain}
\altaffiltext{32}{California Institute of Technology, Pasadena, CA}
\altaffiltext{33}{Dept. of Physics and Astronomy, University of Alabama, Tuscaloosa, AL}
\altaffiltext{34}{Dept. of Physics and Astronomy, State University of New York at Geneseo, Geneseo, NY}
\altaffiltext{35}{Department of Astrophysical Sciences, Princeton University, Princeton, NJ }
\altaffiltext{36}{Observatoire de Gen\`eve, University of Geneva, Geneva,Switzerland}
\altaffiltext{37}{Dept. of Astronomy, Columbia University, New York, NY}
\altaffiltext{38}{INAF -- Osservatorio Astronomico di Bologna, Bologna, Italy}
\altaffiltext{39}{IPAC/CalTech, Pasadena, CA}
\altaffiltext{40}{Dept. of Astronomy, New Mexico State University, Las Cruces, NM}
\altaffiltext{41}{UPMC--CNRS, UMR7095, Institut d'Astrophysique de Paris, Paris, France}
\altaffiltext{*}{Einstein Fellow}

\begin{abstract}
The Legacy ExtraGalactic UV Survey (LEGUS) is a Cycle~21 Treasury program 
on the Hubble Space Telescope, aimed at the investigation of star 
formation and its relation with galactic environment in nearby galaxies,  
from the scales of individual stars to those of $\sim$kpc--size 
clustered structures. Five--band imaging, from the near--ultraviolet 
to the I--band, with the Wide Field Camera 3, plus parallel optical 
imaging with the Advanced Camera for Surveys, is being collected for 
selected pointings of 50 galaxies within the local 12~Mpc.  The filters 
used for the observations with the Wide Field Camera 3  are: F275W($\lambda$2,704\AA), F336W($\lambda$3,355\AA), 
F438W($\lambda$4,325\AA), F555W($\lambda$5,308\AA), and F814W($\lambda$8,024\AA);  the parallel observations 
with the Advanced Camera for Surveys use the filters: F435W($\lambda$4,328\AA), F606W($\lambda$5,921\AA), and F814W($\lambda$8,057\AA). 
The multi--band images are yielding  accurate recent ($\lesssim$50~Myr) star formation histories 
from resolved massive stars and the extinction--corrected 
ages and masses of star clusters and associations. The extensive inventories 
of massive stars and clustered systems will be used to 
investigate the spatial and temporal evolution of star formation within 
galaxies. This will, in turn, inform theories of galaxy evolution and 
improve the understanding of the physical underpinning of the gas-star 
formation relation and the nature of star formation at high 
redshift. This paper describes the survey, its goals and observational strategy, 
and the initial science results. Because LEGUS will provide a reference survey 
and a foundation for future observations with JWST and with ALMA, a large 
number of data products are planned for delivery to the community. 
\end{abstract}

\keywords{galaxies: general -- galaxies: star clusters: general -- galaxies: star formation -- galaxies: stellar content -- ultraviolet: galaxies -- ultraviolet: stars}

\section{Introduction}

Major progress in the characterization of star formation in galaxies, one of the main processes that governs galaxy evolution, has been enabled by decades of observations from the ground in the optical, near--infrared, and mm/radio, and, more recently, from space in the ultraviolet, optical, and 
infrared wavelength range, with facilities such as the 
Hubble Space Telescope (HST), the GALaxy Evolution eXplorer \citep[GALEX,][]{Martin2005},  the Spitzer Space Telescope \citep[SST,][]{Werner2004}, 
the Wide--field Infrared Survey Explorer  \citep[WISE,][]{Wright2010}, and the Herschel  Space Observatory \citep[HSO,][]{Pilbratt2010}. 

The same observations have also highlighted that we are still missing a critical piece in the star formation puzzle. Star formation has been investigated so far 
on two fundamental scales: those of individual stars, stellar clusters and associations on parsec scales, and those of galaxy disks on kpc scales. Vast differences 
in the observational capabilities and observing strategies required to probe pc and kpc scales have caused the work to effectively proceed on parallel and non--intersecting tracks. As a result, we have not yet characterized the links  between the two scales, which represents a major barrier to the development of a 
predictive theory of star formation.

For instance, the tight scaling relation found in galaxies at all redshifts between star formation and the gas reservoir, when these quantities are averaged 
over kpc or larger scales \citep[the Schmidt--Kennicutt Law,][]{Schmidt1959,Kennicutt1998,KennicuttEvans2012,Daddi2010,Genzel2010}, breaks down  when zooming 
into their basic constituents. Young stars and star clusters appear basically uncorrelated with molecular 
clouds over scales smaller than $\sim$100--200~pc \citep{Onodera2010,Momose2010,Schruba2010}. This trend may be due to the onset of 
two non--exclusive effects: (1) the increasing scatter in both tracers of star formation rate and gas clouds, due to small number statistics at small scales 
\citep{Calzetti2012,Kruijssen2014,daSilva2014}; and (2) the finite characteristic timescale of the association between a group/cluster of young stars and its natal 
cloud \citep{Kawamura2009}. Within molecular clouds, the closest association is observed between the dust--enshrouded star formation and the densest gas 
components  \citep{Gao2004,Lada2010,Lada2012}, while star formation scales in  a non--uniform manner with the molecular gas reservoir 
\citep{Heiderman2010,Gutermuth2011}. Thus, the physical underpinning of the  Schmidt--Kennicutt Law lies in the still unknown nature of the 
link between  large--scale and small--scale star formation  \citep[e.g.,][]{Hopkins2013}.  

Star formation has a profound role in the formation of the macroscopic components of galaxies. The giant, kpc--size, $\approx$10$^8$~M$_{\odot}$ clumps observed in star--forming galaxies at redshift z$>$1 \citep{Immeli2004,ElmegreenElmegreen2005,ElmegreenD2007,Elmegreen2009,Genzel2008,Genzel2011,ForsterSchreiber2011} may be the result of gravitational 
instability in gas--rich turbulent disks \citep{Elmegreen2008,Dekel2009,Ceverino2010,Ceverino2012}, fed through accretion of cold gas via smooth flows 
\citep{Keres2005,Dekel2009b,Giavalisco2011}. These giant star--forming clumps are expected to evolve by migrating toward the center of the host galaxy to 
coalesce into the bulge or by being disrupted by tidal forces or feedback to form the thick disk \citep{Bournaud2009,Murray2010,Genel2012}. While this 
scenario is broadly supported by current observations of high redshift galaxies \citep{Elmegreen2009,Guo2012}, an important consideration is that no 
clumps of comparable size and mass are observed in present--day galaxies \citep{ElmegreenD2009}. Local star--forming irregulars, however, do show a 
clumpy structure, akin to that of high--redshift galaxies, but with several ten--fold lower clump masses \citep{ElmegreenD2009,Elmegreen2012}. How  
galaxies form stars over large scales within their bodies has clearly changed with time, likely in response to both internal and external factors, 
but the evolution of this trend has not been mapped, yet; nor have we a full grasp of the link between star forming structures at any scale and the global 
properties of their host galaxies \citep{Dobbs2011,Hopkins2013,Dobbs2013}. 

GALEX has established the local benchmark for the comparative interpretation of the rest-frame UV galaxy substructures of high redshift optical surveys, 
reflecting the evolution in galaxy populations between these epochs \citep[e.g.,][]{Petty2009}. 
The detailed and systematic UV imaging of local galaxies conducted by GALEX  \citep{GildePaz2007} has revealed the morphological diversity of star forming 
environments, from currently bursting regions to fading, intermediate age populations \citep[e.g.,][]{Lee2011,JohnsonBD2013}. Because the UV is produced by 
stars with masses that extend to  significantly lower  values 
than those needed to produce ionizing photons, the UV emission has a roughly ten--fold longer timescale than other star formation tracers such 
as H$\alpha$ \citep[e.g.,][]{KennicuttEvans2012,Calzetti2013}, and can probe regions of very low star formation rate surface density, including the extended UV disks that are still the subject of extensive investigation \citep{Thilker2005,GildePaz2005,Yi2005,Thilker2007,Jeong2007,Dong2008,Efremova2011,Alberts2011,Lemonias2011,Koda2012}.

GALEX has also underscored the complexity of interpreting the UV light as a star formation rate tracer. UV images of nearby galaxies display a mix of discrete clumps 
embedded in a background of apparently diffuse emission. The color and intensity of the diffuse UV emission is highly variable, as a function of position 
within a galaxy and with respect to regions of prominent current star formation. Some of this diffuse emission may originate from aging, dissolved 
stellar clusters and associations \citep[e.g.,][]{Cornett1994, Crocker2014}, but some may also be dust scattered light from neighboring star--forming regions 
\citep{Popescu2005}, although evidence for the latter is being brought into question \citep{Crocker2014}. The star formation history and population mixing play 
important roles in the observed UV colors of a galaxy, both locally and globally \citep{JohnsonBD2013}. Quantifying this role and the parameters governing it 
can inform the strategies for extending the use of UV colors as diagnostics of dust attenuation from starbursts \citep{Calzetti2000} to normal star--forming galaxies
\citep{Hao2011, Boquien2012}, crucial for high redshift galaxy population studies.

Finally, combined observations of nearby galaxies with GALEX, the SST, and the HSO have provided an excellent characterization of the global kpc--scale properties of star formation  spanning the complete range of gas richness, star formation activity, stellar mass, metallicity and morphology in galaxies. GALEX and the SST$+$HSO have provided complementary pictures of star formation across the full disks of galaxies, by probing both the direct UV emission from young, massive stars and the 
dust--reprocessed light from the same stars in the IR. These have been used, among other things,  to define relatively unbiased tracers of star formation rate \citep[SFR; e.g.,][]{Calzetti2005,Kennicutt2009,Lee2009,Liu2011,Hao2011} and investigate the star formation laws and efficiency throughout and across galaxies \citep[e.g.,][among many others]{Kennicutt2007,Leroy2008,Schiminovich2010,Schruba2011,Calzetti2012}. 

Amid the richness of all these datasets, the limited angular resolution of GALEX (5$^{\prime\prime}$, corresponding to $\sim$20 pc even in M33 at 840~kpc 
distance) has prevented linking the large--scale star--forming structures to the physical components (clusters, associations, isolated massive stars) that produce the variety of structures observed in galaxies at all redshifts. The HST Treasury Program LEGUS (Legacy ExtraGalactic UV Survey, GO--13364) 
is designed to bridge the gap between the small--scale and large--scale star formation, by exploiting the combination of  high--angular resolution (about 
70 times that of GALEX) of HST  with the UV capabilities of the imaging cameras aboard the telescope. 

\begin{figure}
\figurenum{1}
\rotate
\plotone{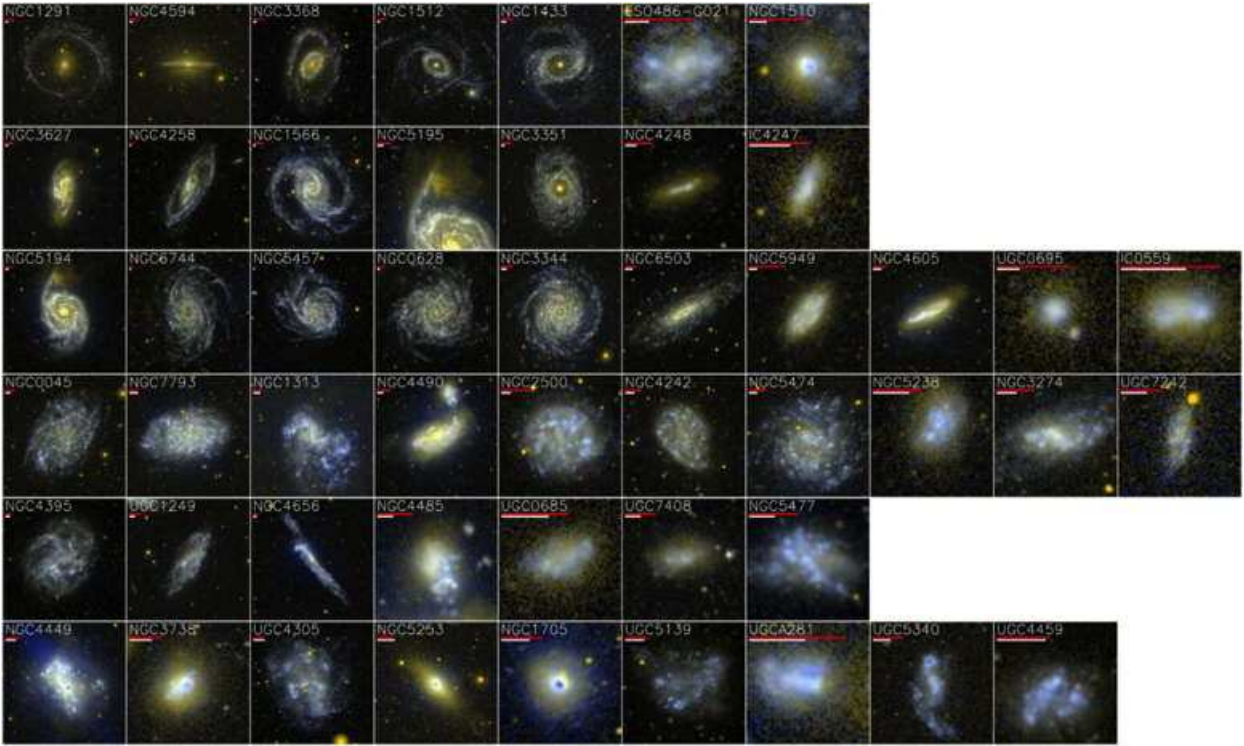}
\caption{Montage of the GALEX 2--color (far--UV and near--UV) images for the 50 LEGUS sample galaxies. The name of each galaxy 
is shown in each panel, together with two bars to provide a scale: 1$^{\prime}$ (red) and 1~kpc (white) in length, respectively. The physical size is calculated using the distances 
listed in Table~\ref{tab1}. Each panel has a size equivalent to 1.5$\times$D$_{25}$. The 50 galaxies are ordered according to morphological T--type, using 
the groupings of Table~\ref{tab1} for each row. Within a row, the galaxies are ordered according to descending stellar mass, and, for ties in stellar mass, according to 
descending HI mass. As some morphological types have large uncertainties (Table~\ref{tab1}), some of the LEGUS galaxies may be mis--classified. Other galaxy properties 
are listed in Table~\ref{tab1}.
\label{fig1}}
\end{figure}

LEGUS is collecting 5--band imaging (NUV,U,B,V,I) of 50 nearby, star--forming galaxies in the distance range $\sim$3.5--12~Mpc, i.e., in the local volume of 
the Universe within which HST can simultaneously resolve and age-date the young stellar populations on pc scales and  probe the galaxies'  structures on 
kpc scales (Figure~\ref{fig1}). The mostly well--known, archetypal galaxies in the LEGUS sample have a large suite of existing multiÐwavelength ancillary 
data  with GALEX, the SST, and other space and ground--based facilities  \citep{Kennicutt2008,Dale2009,Lee2011}, which have been 
used to characterize their large--scale star formation. The sample  covers the full range of galaxy mass, morphology, SFR, sSFR 
(specific SFR=SFR/mass), metallicity, internal structure (rings, bars), and interaction state found in the local volume.  

LEGUS aims at providing complete inventories of young stellar populations and structures, with full characterization
of their ages, masses, extinctions, and spatial distributions,  in order to enable a host of scientific applications, including: quantify how
the clustering of star formation evolves both in space and in time; discriminate among models of star cluster evolution;  investigate the impact of the 
recent star formation history on the UV star formation rate calibrations. In this respect, the UV photometry 
acquired by LEGUS is critical for the age--dating and identification of young massive stars and stellar systems, and the
reconstruction of the recent star formation histories (SFH) over the past $\sim$50~Myr. To achieve these goals, the UV--optical photometric 
observations are designed to provide extinction--free ages, luminosities, sizes, and masses down to $\sim$1--3$\times$10$^3$~M$_{\odot}$ for 
clusters and associations younger than 100~Myr. The UV images are also providing sufficient contrast to isolate, identify, and measure individual stars 
down to 7--10~M$_{\odot}$, in intermediate--to--low density environments. The expectation is to ultimately collect  several hundreds to thousands of clusters, 
stars, and associations per bin of SFR, sSFR, morphological type, and mass. 

In this respect, the LEGUS UV observations are complementary to the current state--of-the--art UV surveys of nearby galaxies 
produced by GALEX \citep[e.g.,][]{GildePaz2007}. With a $\sim$1--degree--square field of view, GALEX has imaged the full disks of nearby galaxies, with enough 
sensitivity to detect their faint outskirts up to galactocentric distances $\sim$2--4~R$_{25}$ \citep{Thilker2007,Lemonias2011}, but at low angular 
resolution ($\sim$5$^{\prime\prime}$). Conversely, the HST observations that LEGUS is obtaining have sufficiently high angular resolution 
($\sim$0.07--0.08$^{\prime\prime}$) to resolve individual star clusters and bright stars within galaxies, but typically cover only a fraction of a galaxy's body, 
i.e., regions $\sim$3--9~kpc in size.  GALEX has imaged the galaxies in two UV filters, centered at 1,524~\AA\ and 2,297~\AA, respectively, while LEGUS 
imaging spans the galaxies' NUV--to--I spectral energy distributions, from 2,700~\AA\ to 8,000~\AA.

A prior HST program, the Wide Field Camera 3 Early Release Science (WFC3--ERS, GO--11360; P.I.: O'Connell) has paved the road for LEGUS by observing 
a few nearby galaxies with a similar  filter set. The WFC3--ERS observations have been used for a wide range of scientific investigations, including studies 
of young star cluster populations \citep{Chandar2010b,Whitmore2011,Chandar2014}, of the high--mass stellar initial mass function 
\citep{Andrews2013,Andrews2014}, of the ages and metallicities of a globular cluster population \citep{Kaviraj2012}, and of the 
recent--past star formation history of an early--type galaxy \citep{Crockett2011}.

This paper is organized as follows: Section~2 describes the specific scientific goals of LEGUS; Section~3 provides details on the sample selection; 
Section~4 describes the observations and the higher level data products, all of which will be delivered to the community; a few of the initial science 
results are given in Section~5, while Section~6 describes the Public Outreach initiative of the project. A summary is provided in Section~7. 

\section{Scientific Objectives}

The science goals described in this Section are a small sub--sample of the multiple applications  that a diverse, multi--wavelength 
survey like LEGUS can enable. 

\subsection{The Hierarchy of Star Formation}

The mechanisms that govern and regulate star formation over kiloparsec scales in galaxies  are an unsettled issue. According to one 
scenario,  stars are born either in clustered environments or in diffuse, low-density environments, and the field and cluster environments 
have physically distinct modes of star formation  \citep[e.g.,][]{Meurer1995}. In the two environments, stars may 
even be characterized by different stellar Initial Mass Functions \citep[IMFs; e.g.,][]{Massey1995,Lamb2013}. The opposite 
scenario, that all stars form in clusters, has found more traction in recent years in its `weaker' formulation, i.e., that 
all stars form in structures that  are a continuous, scale--free 
hierarchy from parsecs to kiloparsecs  \citep[]{Lada2003, Zhang2001, Elmegreen2003, Gutermuth2005, Sanchez2010,Gouliermis2010, 
Bressert2010}. Bound star clusters occupy the densest regions of the hierarchy \citep[e.g.,][]{Elmegreen2010}, but most of the structures 
are unbound, and their stars disperse over time, forming the field population. 

The evolution of the unbound structures with time is the subject of intense recent investigation. The 
erasure of structures occurs on time scales of $\sim$100--200~Myr in the LMC \citep{Bastian2009,Baumgardt2013} and SMC 
\citep{Gieles2008}, consistent with the dynamical crossing times for those galaxies. In M\,31, clustered stellar structures survive for 
a longer period of time, $\sim$300~Myr \citep{Gouliermis2014}. 
Age--dependent clustering is observed in other galaxies:  M\,51 \citep{Scheepmaker2009}, and  
NGC\,1313 and IC\,2574 \citep{Pellerin2007, Pellerin2012}, with lower mass stars showing progressively weaker 
clustering. In starburst galaxies, the UV--bright stellar populations outside of star clusters lack the early--type stars (earlier than B)
that dominate the young star clusters; this is indicative of either dispersal of structures over shorter timescales than the Magellanic Clouds 
($\sim$10~Myr instead of $\sim$100~Myr) or evidence for a steeper IMF in regions outside the cluster locales 
\citep{Tremonti2001,Chandar2005}. \citet{PortegiesZwart2010} find that at young ages ($\lesssim$10~Myr) stars 
are distributed in a continuum of structures, but a bimodal distribution of bound clusters and diffuse population 
develops at later ages. A more complex picture may ultimately emerge, if both bound and diffuse structures 
co--exist  in systems that are younger than $\sim$5~Myr \citep{Gouliermis2014b}. 

The nature and the spatial and temporal evolution of the hierarchical structures of stars can also constrain
models of massive star formation \citep[see][for a review]{Tan2014}. Core collapse models \citep[e.g.,][]{Krumholz2009} 
allow for occasional isolated stars without associated clusters, implying that  scale--free hierarchies can be constructed also by  
individual massive stars.  Conversely, competitive accretion models \citep[e.g.,][]{BonnellBate2006} require that O stars 
always form in clusters more massive than the star itself, and are thus generally located at the peak densities of the hierarchy, 
with the exception of runaway stars \citep{deWit2005, Gvaramadze2012}.

The information that has been gathered so far is sparse, owing mainly to the absence of systematic high angular resolution, 
multi--band surveys that can separate and classify stars and structures as a function of age, in a variety of environments, 
including moderately crowded ones.

By collecting large and coherently measured samples of clusters and massive stars with well characterized ages and 
masses for a variety of galactic environments, LEGUS will enable a quantitative picture of the clustering of star formation, 
by addressing: (1) whether the hierarchy has characteristic scale(s); (2) how the hierarchy evolves with time; (3) whether 
its characteristics and evolution depend on the environment. 

Tools that will be employed to address the three points above include well--established techniques, such as friends-of-friends algorithms, 
minimum spanning trees, and angular two--point correlation functions: 
\begin{equation}
w(\theta) = A \theta^{(1-\alpha)},
\end{equation}
where the amplitude $A$ is related to the correlation length of the clustering, and $\alpha$ 
measures the strength of the clustering \citep[e.g.,][]{Peebles1980}. By definition, the two--point angular correlation function quantifies the excess 
probability above a random distribution of finding one object (e.g., a star) within a specified angle $\theta$ of another object. If the clustered distribution 
is self--similar (scale--free), $\alpha$ is related to the correlation dimension D$_2$ via: $\alpha$=3$-$D$_2$ \citep{Heck1991}. If the 
correlation length evolves with the age of structures, it will provide clues on, and help quantify, the clustering dispersal timescale.
These tools will be applied to clusters, massive stars, and associations to identify 
common age stellar structures, and to derive the correlation length as a function of age and location within galaxies. 

The clustering statistics will be studied as a function of the kpc-scale properties, both galaxy-wide (SFR, sSFR, morphology, mass, interaction state) 
and local (SFR/area, galactocentric distance, presence of structures like arms, bars, rings).  In a recent analysis of twelve LEGUS galaxies, \citet{Elmegreen2014} 
conclude that clustering of star formation remains scale--free up to the largest scales observable, for both starburst galaxies and  galaxies with more quiescent levels of star formation. This suggests that hierarchically structured star--forming regions several hundred parsecs or larger represent common unit structures, and is 
consistent with a picture in which star formation is regulated by turbulent processes. These conclusions will be generalized by a more extensive 
investigation of the LEGUS galaxies.  Where data are available,  cross--correlation analyses can be 
expanded to be between clusters/stars and features in the ISM, as traced by ground or space-based
H$\alpha$, HI, CO (e.g. ALMA), and far--infrared (Spitzer, Herschel) emission. 
The cross--correlation lengths between stellar and interstellar tracers, can then be used to 
test theories of cloud formation \citep[e.g.,][]{Elmegreen2006}. Predictions from dynamical models 
\citep[e.g.,][]{Dobbs2010,Dobbs2010b} can be compared with observations on clustering, and increase the predictive power of models for 
higher redshift galaxies. The ultimate goal of this part of the analysis is to establish whether the young ($\lesssim$100--200~Myr) 
field population results entirely from the dissolution of clustered star formation that originated elsewhere, or requires a component 
of in--situ star formation.

\subsection{The Evolution of Stellar Clusters}

Star clusters face a number of challenges to their survival: most are born unbound (as discussed in the previous section), but many 
also become unbound as their stellar populations evolve. Between 70\% and 98\%  of stars born in star clusters disperse within the first 
10--20~Myr as an effect of the rapid gas expulsion phase driven by massive star winds and supernova explosions 
\citep[`infant mortality', e.g.,][]{Lada2003,Fall2005,GoodwinBastian2006,GielesBastian2008}. The subsequent evolution of star clusters 
depends on a number of factors; additional disrupting mechanisms include  mass loss due to stellar evolution, 
stellar escape due to two--body relaxation, and the tidal field of the host galaxy. The timescales over which each mechanism 
dominates may or may not differ, depending on model assumptions \citep{Fall2009,Gieles2011}. 

Despite these challenges, the large numbers of young compact clusters present in actively star-forming galaxies has led to the suggestion 
that they could represent the present--day analogs of globular clusters \citep[e.g.,][]{Zepf1999,Whitmore2003}. The question of whether, in what 
fraction, and under which conditions these young clusters can survive for $\sim 10$ Gyr is still highly controversial because of the lack of consensus on 
the influence of the environment on their evolution.  While the `infant mortality' phase of gas removal is probably mass-independent, there is not yet an 
agreement on whether the later phases are mass--dependent and more massive clusters live longer. In the `Mass--Dependent--Disruption' scenario   
\citep[MDD, e.g.,][]{Lamers2005, Bastian2012}, a cluster lifetime depends on its mass, as $\tau\propto$M$^{\gamma}$, with $\gamma\sim$0.62, as derived 
from models; in addition, clusters in weak tidal fields or with few interactions with surrounding molecular clouds have longer lifetimes. In the `Mass--Independent--Disruption'  
scenario \citep[MID, e.g.,][]{Whitmore2007, Fall2009, Chandar2010}, the evolution of a star cluster is independent of its mass and the ambient conditions, 
leading to a universal expression for the number of star clusters present at any given time and mass range, [d$^2$N/(d$\tau$ d$M$)]$\propto\tau^{-1}$~M$^{-2}$, with roughly 90\% of the clusters disrupting in each decade of time \citep[e.g.,][]{Chandar2010b,Baumgardt2013}. 
The two scenarios do not necessarily need to be mutually exclusive: the MID may result from MDD in a hierarchical interstellar medium, 
although the ambient conditions will still have strong influence on the outcome \cite[e.g.,][]{ElmegreenHunter2010}.  For instance, \citet{SilvaVilla2014} find evidence for environmental dependence 
in the cluster population of the nearby spiral M\,83, with  a high disruption rate toward the center and little or no disruption in the outer regions. Conversely, the cluster population in M\,31 does not show evidence 
for disruption over the first 100~Myr, but only for older ages \citep{Fouesneau2014}. A spin--off of the two scenarios is whether 
the maximum cluster mass  observed in each galaxy is a size--of--sample effect \citep[e.g.,][]{Mora2009, Whitmore2010} or a physical truncation  \citep[e.g.,][]{Gieles2009,Bastian2012b}. A truncation is observed in giant molecular clouds in the Milky Way and in nearby galaxies  \citep[e.g.,][]{Rosolowsky2005} and is thought to be related  to the Jeans mass in the galactic disk \citep[e.g.,][]{KimOstriker2006}. 

The main obstacle to discriminating between MDD and MID is the lack of large and homogeneously selected samples of star clusters with masses below $\sim$10$^4$~M$_{\odot}$, where the effects of disruption, especially if mass--dependent, would be most evident. 

LEGUS will offset this limitation by providing an order--of--magnitude larger sample of galaxies with well--characterized cluster populations than currently available. The extensively--tested univariate mass and age distributions (dN/dM--vs--M and dN/d$\tau$--vs--$\tau$) at constant age and mass, respectively, and the bivariate (d$^2$N/dM\,d$\tau$) function \citep{FallChandar2012} will be measured for extensive star cluster populations down to $\sim$a few 1,000 M$_{\odot}$, across the full range of local galactic environments. Other methods involving the short--scale ($\lesssim$100~Myr) time evolution of the blue colors of the cluster populations  
as predicted by the two different scenarios, MDD and MID, are also being investigated. 

As well as discriminating star cluster evolution scenarios, LEGUS will address a number of long-standing questions on the role of star clusters in star formation processes. By relating cluster formation and cluster properties to the star formation rates and morphologies of their host galaxies, the following 
can be addressed: 1) constrain the fraction of stars that form in clusters and search for environmental dependencies; 2) study the cluster luminosity/mass function and determine if a characteristic mass exists in the distribution; 3) measure the size (radius) distribution of the clusters and determine if this has a dependence on environment; and 4) determine the cluster formation histories of these galaxies over the past $\sim$Gyr. 

\subsubsection{Testing Bar Evolution with Star Clusters}

Homogeneous samples of star clusters can be used to test models of sub--galactic structure evolution, as young clusters trace the underlying GMC distributions 
within galaxies. Inner and outer stellar rings and spiral arms are sites of active star--formation with large concentrations of gas \citep{ButaCombes1996}, similar to 
what is found in  bars \citep{Sheth2005}. Strong bars are expected to drive gas into the centers and fuel an AGN, implying that the gas loses  a factor 
$\sim$10$^4$ in angular momentum. One of the extant questions is whether the higher star formation in these regions is due to triggering by density waves or 
to more/larger molecular clouds \citep[GMCs; e.g.,][]{Nimori2013}. This can be tested, as the  scales and angular momenta of GMCs impact the mass and age 
distributions of stellar clusters in these regions. The location, distribution, and radial trends of the star cluster populations in the LEGUS sample can be compared 
with the large scale features and surface mass over--densities, as derived from Spitzer 3.6 and 4.5~$\mu$m maps \citep{Meidt2012}, to help constrain dynamical 
models for the formation of different structures and possible mechanisms for gas accretion and inflow. Future ALMA observations will  map the location, size, and 
distribution of the GMCs, thus providing a direct comparison with the young star cluster populations derived in this project.

\subsection{UV SFR Calibrations and the Recent Star Formation History}

Well calibrated and accurate SFR indicators are necessary to bridge our understanding of resolved stellar populations in galaxies in the local universe with their unresolved counterparts at high redshift \citep{KennicuttEvans2012}. The extinction--corrected UV indicator is one of the most commonly used  SFR indicators \citep[e.g.,][]{Kennicutt1998b, Salim2007}. Recent studies have highlighted potentially significant discrepancies between standard calibrations and the newest star/stellar population
 evolution models. Evolutionary models that include stellar rotation result in a 30\% smaller UV--to--SFR calibration \citep{Levesque2012}, producing a factor 2 discrepancy with SFRs derived from core collapse supernovae \citep[CCSNe, ][]{Smartt2009, Horiuchi2013}. UV--based SFRs can also be affected by 
 environment--dependent IMF variations at the high--mass end \citep{Lamb2013}. These comparisons, however, are sensitive and degenerate with variations of the SFH over the most recent 50--100~Myr, where$\sim$80\% of the UV emission is produced. 
 
 Post--starburst conditions in luminous galaxies and sporadic star formation in faint, low--mass galaxies \citep[e.g.,][]{JohnsonBD2013} can yield UV--based SFRs that 
 are discrepant by factors of a few with those derived at other  wavelengths (e.g., H$\alpha$), mainly owing to the different timescales involved by the different emission processes, i.e., $\gtrsim$100~Myr for the stellar continuum, non--ionizing UV and $\approx$10~Myr for the hydrogen recombination line H$\alpha$ 
 \citep[e.g.,][]{KennicuttEvans2012,Calzetti2013}. Some of these systematic effects may be at the basis of (at least part of) the observed trend for 
 decreasing H$\alpha$/UV ratio in 
 increasingly fainter dwarf galaxies \citep{Lee2009,Fumagalli2011,FicutVicas2014}. Variations in SFHs are as viable an explanation as the systematic changes in the 
 high--end of the IMF invoked to account for the H$\alpha$/UV trend \citep{Weisz2012}. Again, accurate measures of recent SFHs are key for addressing these issues.

One of the goals of LEGUS is to enable accurate ($\delta(age)/age\approx$10\%--20\%) determinations of SFHs in its sample galaxies. The LEGUS 
UV observations resolve the majority of the stars above $\approx$7--10~M$_{\odot}$ at all distances, in the disks and 
in sparse groups and OB associations; the outer regions of star clusters can be partially resolved in the closest ($<$5--6~Mpc) galaxies.  In color--magnitude diagrams (CMDs) that involve a UV band, the NUV images separate Main Sequence (MS) stars from the Blue Loop (BL) core helium burning giants, in a cleaner sequence than optical CMDs where BL and MS stars overlap 
\citep[][ also Section~6]{Tolstoy1998}.
The luminosity of the BL stars depends mainly on their mass and, therefore, fades monotonically with age. With a clear MS-BL separation, the nearly one--to--one correspondence of luminosity and age for BL stars can be leveraged to convert directly the luminosity function into the SFH \citep{DohmPalmer1997,Tolstoy2009}. 

Population synthesis techniques \citep{Tosi1991,CignoniTosi2010}, based on comparing observed and synthetic (Monte Carlo--based) CMDs, can then be employed to derive the detailed SFHs, after applying  star--by--star extinction corrections to the data \citep{Kim2012}. Assumptions can be included in the models for a variety of star formation laws, IMFs, binary fractions, age-metallicity relations, and stellar evolution models to test their impact. 
This method has been already applied to galaxies at distances from $<$1~Mpc to 18 Mpc, i.e., from the Local Group to IZw18  \citep{Cignoni2011, Cignoni2012, Annibali2013}. 

CMDs of resolved stars and Bayesian techniques applied to partially resolved clusters, sparse groups and OB associations \citep[e.g.,][]{Weisz2013} will also place constraints on the high--end of the stellar IMF, above 7--10~M$_{\odot}$. When combined with the local SFHs in the field and in sparse groups and OB associations, the UV emission will be mapped back to the SFRs, and solve or set limits on the discrepancies with the CCSNe.  

\subsection{Multiple Stellar Populations in Globular Clusters}

One of the most exciting discoveries in recent stellar populations research is the presence of complex populations in massive globular clusters (GCs).  The most prominent examples of this phenomenon are the multiple main sequences in $\omega$~Cen \citep{Anderson1998} and the triple main sequence in NGC\,2808 \citep{DAntona2005}.  These GCs apparently contain a significant ($\sim 20\%$) population of He-rich ($Y \sim 0.4$) stars \citep{Piotto2005} that likely formed in a second stellar generation.  For a population at a given age and metallicity, the main-sequence turnoff mass decreases with increasing $Y$, such that He enhancement has a significant effect on the later evolutionary phases. Most notably, the temperature 
distribution of horizontal branch (HB) stars becomes hotter at increasing $Y$ \citep[e.g.,][]{Chung2011}, and thus the same massive GCs exhibiting multiple main sequence also host HB stars extending to extremely high effective temperatures \citep[$T_{\rm eff} > $~25,000~K;][]{DAntona2002,Brown2010}. At distances greater than 4~Mpc, individual stars in globular clusters cannot be resolved, but a strong UV excess in integrated light is a powerful diagnostic for those massive GCs hosting multiple populations \citep{Kaviraj2007,Mieske2008}.  

Although massive GCs can exhibit an extended HB morphology over a wide range of metallicities, for old GCs at typical masses, HB morphology is correlated with metallicity.  The HB stars in low-metallicity GCs tend to be bluer than the RR Lyrae gap, where they will dominate the UV light output \citep[e.g.,][]{Ferraro1997}.  The HB stars in high-metallicity GCs tend to fall in the red clump, and in such GCs the UV light may be dominated by the hottest blue straggler stars \citep[e.g.,][]{Ferraro2001}.

LEGUS enables tracing the presence of the UV--bright clusters in a wide variety of galactic environments, thus providing statistics on their frequency and complementing another Cycle~21 Treasury program, which will provide UV photometry of Galactic GCs (PI Piotto, GO-13297). 

\subsection{The Progenitors of Core--Collapse Supernovae}

All stars with mass above $\sim$8~M$_{\odot}$  explode as supernovae at the end of their lives \citep[although see, e.g.,][]{Kochanek2008,Kochanek2013}. Core--Collapse Supernovae (CCSNe) counts provide a sanity check for stellar--emission--based SFR indicators (section~2.3). Identifying and investigating the nature of the progenitors of CCSNe impacts many areas of astrophysics: stellar evolution, gamma ray bursts, the origin of the chemical elements and the evolution of galaxies. Progenitors of SNe Ib/Ic ($\sim$1/4 of all CCSNe) have so far eluded detection in HST optical imaging \citep[although see][]{Cao2013}: their T$_{eff}$ are large ($\log T_{eff}({\rm K})  \approx 5.3$), a consequence of either stripping by strong line-driven winds from a single Wolf-Rayet (WR) star or mass exchange in an interacting binary system \citep{Yoon2012, Eldridge2013}. Both models lead to a hot, luminous progenitor, best detected in the UV. Additionally, some SNe IIn appear to be associated with luminous, blue objects, possibly LBVs \citep[e.g.,][]{GalYam2009,Smith2011,Ofek2014}. LEGUS is providing the `pre-CCSNe' UV images that can be used in the future to identify progenitors of CCSNe. Based on current statistics, the archival images provided by this project will enable the detection of up to 12 progenitors for the CCSNe that are expected to explode in the galaxies over the next 10 years, nearly tripling the existing numbers\footnote{While this paper was being written, a SN II, 2014bc, was discovered in the southern pointing of NGC\,4258 \citep{Smartt2014}. LEGUS UV, U, B images obtained a little over one month before the SN's explosion are available, together with archival V,I images \citep{VanDyk2014}.}. This can potentially include a nearby SN Ib/c progenitor detection. The existence of multi--color imaging  will enable us to determine a reliable mass function for the CCSNe precursors and to test if the lack of high-mass progenitors is a real effect \citep{Smartt2009}. The same images can be used to study the environments surrounding CCSNe \citep[e.g.,][]{Murphy2011}, look for light echoes around the CCSNe \citep[e.g.,][]{VanDyk2013}, and investigate dust production in CCSNe \citep[e.g.,][]{Sugeman2006}.

\section{Sample Selection}

The science goals briefly presented in the previous section are the drivers upon which the criteria for selecting the LEGUS sample were built. 

The 50 LEGUS targets were selected from the $\sim$400 star--forming galaxies (out of a total of $\sim$470) in the 11HUGS catalog, which has well--defined completeness properties and is limited within $\approx$11~Mpc \citep{Kennicutt2008,Lee2011}. Use of this catalog as a source of nearby targets offers the advantage that extensive ancillary data are already available in public archives  (the NASA Mikulski Archive for  Space Telescopes [MAST], and the NASA InfraRed Science Archive [IRSA]) and in the NASA Extragalactic Database (NED), which enable leveraging the previous characterization of kpc--scale star formation. The ancillary data include the GALEX far-UV and near-UV images, centered at 0.153~$\mu$m and 0.231~$\mu$m, respectively \citep{Lee2011}; ground--based H$\alpha+$[NII] images, and lists of [NII]/H$\alpha$ ratios \citep[][see this paper also for a detailed discussion on the sources of [NII/H$\alpha$ ratios]{Kennicutt2008}; and, for a subsample of 260 11HUGS galaxies, SST IRAC and MIPS image mosaics in the wavelength range 3.6--160~$\mu$m \citep{Dale2009}. These are accompanied by the mid--infrared (3.4, 4.6, 12, and 22~$\mu$m) imaging coverage  by the WISE satellite, also available at IRSA. 

Access to the intermediate scale of star formation is accomplished by limiting the distance range to 3.5--12~Mpc, as a compromise between FoV, spatial resolution, and sampling volume. In this range the 2$^{\prime}$.7 FoV of the UV--optical channel in the Wide Field Camera 3 (WFC3/UVIS) subtends 2.8--9.5~kpc, in most cases a significant fraction of a galaxy's disk, which increases observing efficiency.  In the same distance range, the WFC3/UVIS Point Spread Function (PSF)  FWHM subtends 1--4~pc. Star clusters have sizes between one and several pc \citep{PortegiesZwart2010}, and they are generally resolved at all distances in the LEGUS sample \citep[e.g.,][]{Chandar2011}. For stellar sources, our data yield that MS stars are detected down to 6~M$_{\odot}$ 
in the LEGUS galaxy NGC\,6503, located at 5.3~Mpc distance (Section~6.1). NGC\,6503 has a projected SFR/area$\approx$10$^{-2}$~M$_{\odot}$~kpc$^{-2}$ in 
the region targeted by our observations. From our estimates, we infer that a full census of MS stars down to 7~M$_{\odot}$ at $\sim$6~Mpc ($>$10~M$_{\odot}$ at 
11~Mpc) will be routinely obtained in less crowded regions, such as those typical of the SMC bar, which has $\sim$10$^{-4}$~M$_{\odot}$~kpc$^{-2}$ (about two 
orders of magnitude lower than NGC\,6503), 

Additional conditions imposed on the sample were:  (1) Galactic latitude $\ge$20$^o$, to minimize effects of foreground extinction by our own Milky Way; and (2) inclination  less than 70$^o$ to minimize the dust attenuation along the line of sight and maximize the benefits of the UV observations. The list of galaxies and their principal characteristics in the LEGUS sample are listed in Table~\ref{tab1}.  

The total number of galaxies in the sample ultimately depended on  the requirement that all science goals described in Section~2 be accomplished. The tightest constraints are imposed by the low--mass cluster statistics. In order to characterize the systematics of the cluster mass and age distributions, an accuracy of $\lesssim$15\%--20\% on count statistics in  the 3--10$\times$10$^3$~M$_{\odot}$ mass bin needs to be achieved, per decade of log(age) in the  3--500~Myr range \citep{Bastian2012}. This translates into a sample size of $\sim$500--700 clusters at all ages and masses per object/bin, which are obtained with 1--3 WFC3 pointings  for galaxies with SFR$>$1~M$_{\odot}$~yr$^{-1}$ \citep{Chandar2010} and in $\sim$10 pointings at lower SFRs, the latter thus requiring stacking. We limited our selection to log(SFR)$\gtrsim-$2.3, below which galaxies contain too few massive stars and star clusters ($<$10--20 clusters per galaxy) for accomplishing the science goals described in the previous Section. 

\begin{figure}
\figurenum{2}
\plottwo{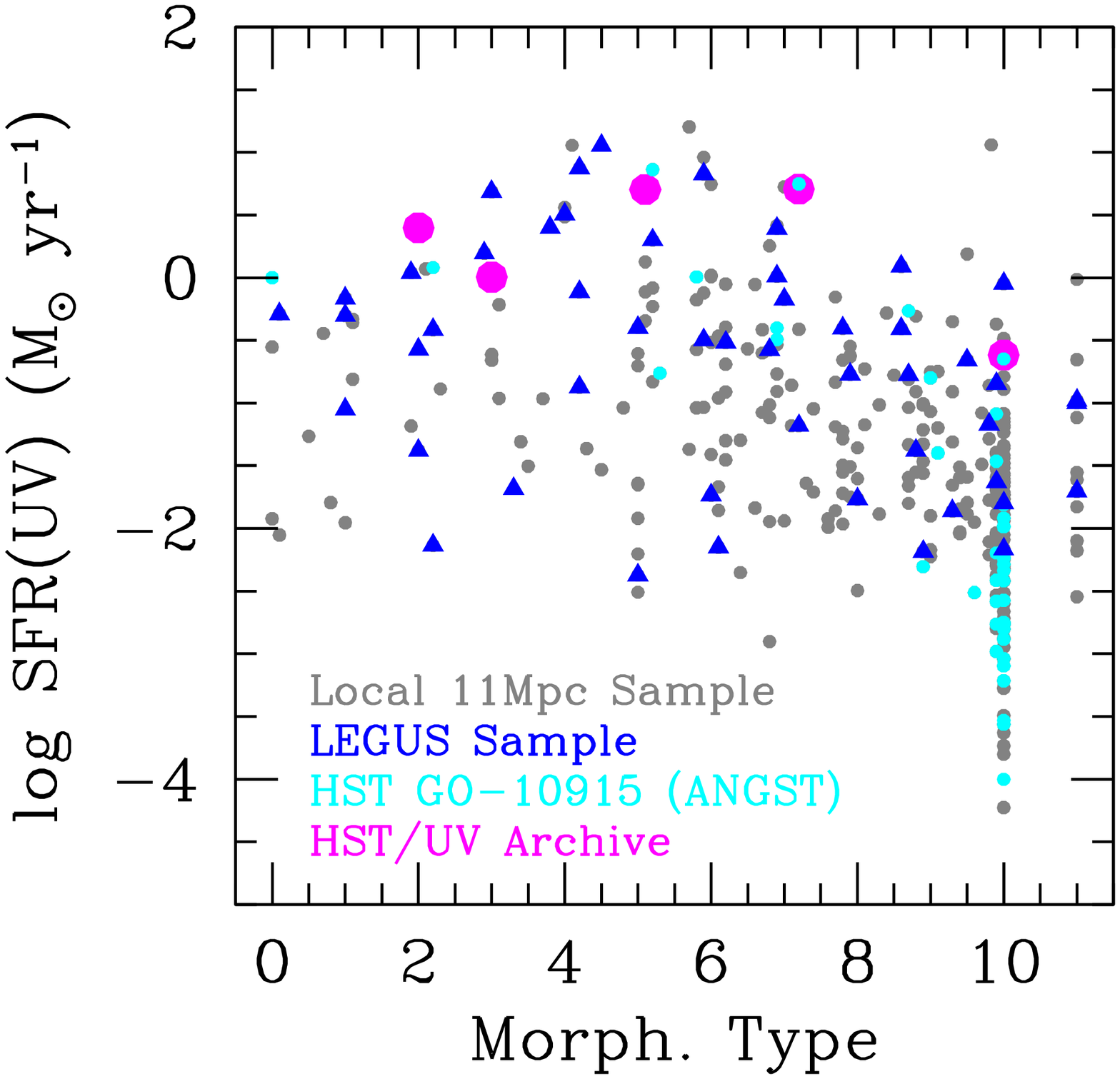}{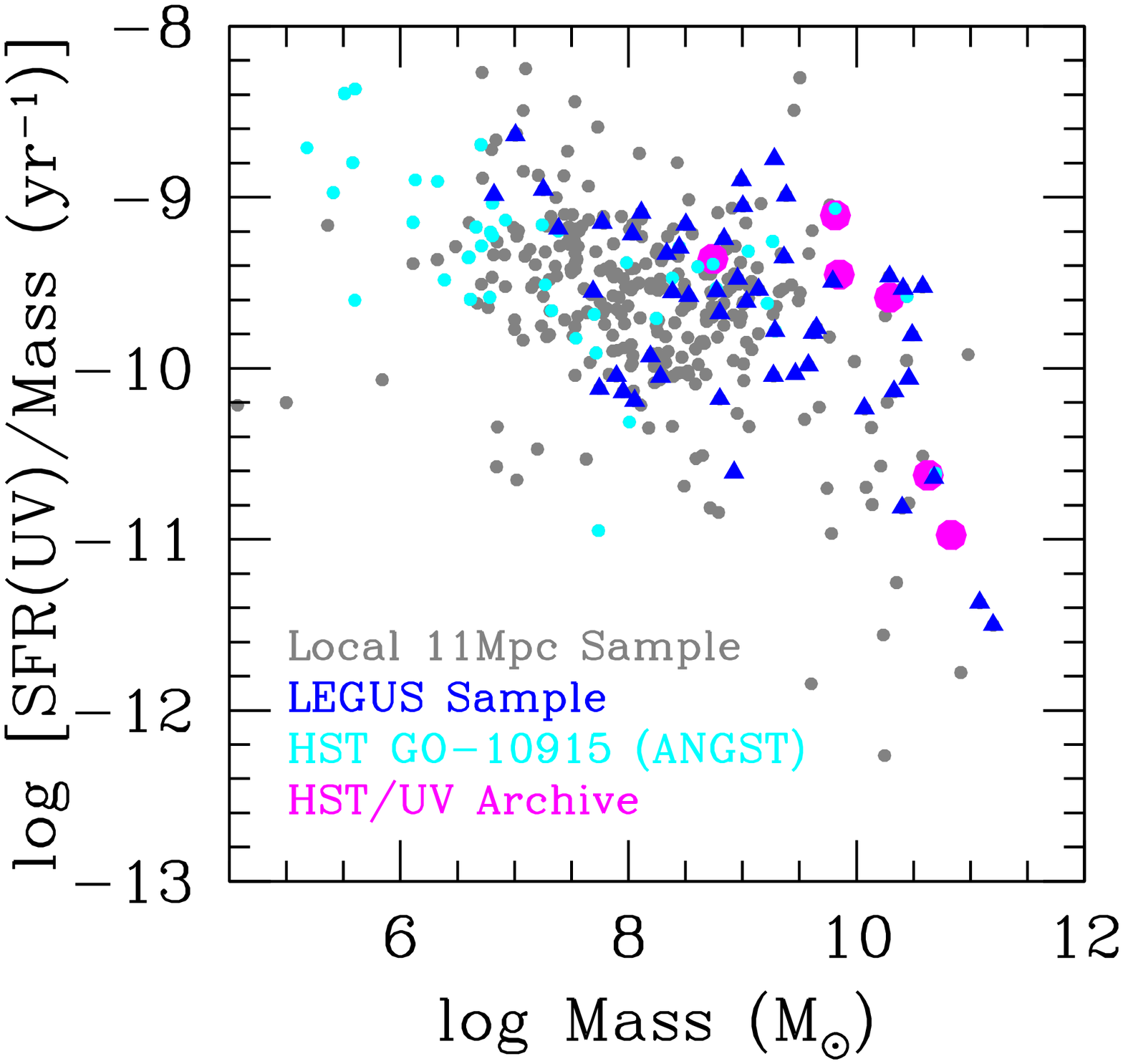}
\caption{Combinations of SFR, sSFR, morphological type, and stellar mass for: the 470 galaxies within 11~Mpc (grey circles) and the LEGUS sample of 50 galaxies (blue triangles). For comparison, also shown is the parameter coverage of the galaxies in the HST GO-10915 program \citep[ANGST,][cyan circles]{Dalcanton2009} and of the galaxies with WFC3/UV archival data (magenta circles; including M31; one galaxy, NGC5128, has T=$-$2.2, and does not appear in the panel to the left). The LEGUS sample covers the full parameter range of local star--forming galaxy properties, except for the lowest mass bin, which is already well represented in previous HST programs. 
\label{fig2}}
\end{figure}

Within the above constraints, the LEGUS sample spans the full range of local galactic properties and environments by populating as evenly as possible bins in the minimal 3 parameters of SFR, sSFR, and morphological type. Each parameter was divided as follows: 3 bins in log(SFR), in the range between $-$2.3 and 1.3 (this being the maximum value observed in the sample), 2 bins in log(sSFR), in the range between $-$11.5 and $-$8.5, and 6 bins in morphology, which include the major morphological types: Sa, Sb, Sc, Sd, Sm, Irr.  Within each bin, 1--2 galaxies were drawn in order to include a range in internal structure (presence/absence of rings and bars) and interaction state (at least 6 interacting pairs are included). The sample size was then adjusted to account for the presence of 6 galaxies already in the HST archive with the pre-requisite wide--field, multi--band photometry at a depth comparable to the LEGUS one, bringing the final LEGUS sample requirements close 
to 50 targets. In down--selecting the specific galaxies to include in the sample, preference was given to 
targets that had one or more of the following properties (in order): (a) archival ACS and/or WFC3 data (typically ACS V and I), with depths comparable to those of the LEGUS observations (see next Section); (b) HI measurements in the literature; (c) oxygen abundance measurements in the literature.  Some of this information is also listed in Table~\ref{tab1}.  Figure~\ref{fig2} shows the distribution of the LEGUS galaxies in the three--parameter space of SFR, sSFR, and morphological type, plus stellar mass. 

The final sample of 50 galaxies (56 when including the 6 galaxies already in the archive: NGC\,224=M\,31, NGC\,2841, NGC\,3034=M\,82, NGC\,4214, NGC\,5128, and  NGC\,5236=M\,83) spans factors of $\sim$10$^3$ in both SFR and sSFR, $\sim$10$^4$ in stellar mass ($\sim$10$^7$--10$^{11}$~M$_{\odot}$; smaller masses are well represented in other HST programs, see Figure~\ref{fig2}), $\sim$10$^2$ in oxygen abundance (12$+$log(O/H)=7.2--9.2). The absolute B magnitude of the galaxies covers the range from $-$13.1 to $-$21.0. All 50 galaxies have ancillary GALEX, SST IRAC$+$MIPS\footnote{NGC\,1433, NGC\,1566, and NGC\,6744 have been observed in all IRAC bands, but in MIPS/24~$\micron$ only.}, and WISE imaging;  41/50 also have ground--based 
H$\alpha+$[NII] imaging from either 11HUGS or the SINGS project \citep{Kennicutt2003b}. These lower--resolution ancillary data trace the large scale star 
formation and galactic environment that this project plans to link  to the small and intermediate scale star formation probed by the HST data. Many of the galaxies in LEGUS are iconic objects (e.g., from the Messier Catalog), with extensive ancillary data that go well beyond those listed here, which maximizes their legacy value. 


\section{Observations}

About half of the galaxies in the LEGUS sample are compact enough that one pointing with the HST WFC3/UVIS will encompass the entire galaxy or most of it  out to 
a UV surface brightness of 3.5$\times$10$^{-19}$~erg~s$^{-1}$~cm$^{-2}$~\AA$^{-1}$~arcsec$^{-2}$, as determined  from the GALEX images. This corresponds to  
m$_{AB}$(NUV)$\approx$27 mag~arcsec$^{-2}$, located at $\approx$2/3 of R$_{25}$ \citep{GildePaz2007}. For those cases in which  the galaxy is slightly more extended than the WFC3 FoV, the pointing was chosen to overlap as much as possible with the archival images, when present; in the absence of constraints from archival data, the pointing was positioned to include the center and as much of the outskirts as possible. Of the 50 galaxies, 11 are significantly more extended than the WFC3 FoV. For 9 of these galaxies (see Table~\ref{tab2}) 
multiple adjacent WFC3 pointings were adopted,  generally along a radial direction from the center outward, in order to encompass as many of the star forming regions as possible, and span a range of environments.  As these galaxies tend to be well--known objects, usually with pre--existing wide--field optical images or 
mosaics in the HST archive, our pointings overlap with and complement the archival ones. NGC\,5194 represents an exception to the `radial' strip criterion: in this case, the location of the pointings was chosen to complement same--Cycle GO pointings in the same or similar filters as those used by our project, in order to maximize 
the legacy value of the datasets, while still maintaining overlap with the optical mosaic obtained with the Advanced Camera for Surveys (ACS) in previous Cycles. The two remaining extended galaxies, NGC\,1291 and NGC\,4594, were observed with only one pointing; these are early--type spirals with lower SFR/area than other galaxies, as determined in the UV from GALEX imaging; the single pointing for each galaxy was chosen to be located on known areas of star formation (outside of the central regions), while 
overlapping with pre--existing optical images with ACS. Table~\ref{tab2} lists for each galaxy the new observations, the number of pointings, and the archival images leveraged for this project. As already mentioned above, archival wide--field images are usually from the ACS with V and/or I filters; existing images were deemed acceptable for this project if the exposure time in each filter was at least 700~s, obtained in a minimum of two frames. Figures~\ref{fig3a}--\ref{fig3f} show the footprints of the new and (where appropriate) archival pointings, in addition to the parallel pointings (see below). The 50 LEGUS galaxies were covered in a total of 63 separate pointings. 

In addition to the WFC3 primary observations, parallel observations with the ACS (listed in Table~\ref{tab2}) are also being obtained. The parallel frames generally 
 target the halo/outer regions of the galaxies. The main goal is to obtain distances for some of the galaxies from the Tip of the Red Giant Branch (TRGB), since not  
all galaxies in the LEGUS sample have secure distances based on either Cepheids or the TRGB method (see Table~\ref{tab1}). However, the ACS parallel pointings  were also left, for the most part, basically unconstrained or only moderately constrained, to ensure a high observing efficiency for the program, at the expense of the 
optimization of the parallel pointings. Efficient scheduling has enabled obtaining the UV images for this program early enough in the Cycle to minimize the effects 
of the CTE degradation of the WFC3 UVIS camera.

We required that each pointing be covered by a minimum of 5 broad band filters: NUV (F275W), U (F336W), B (F438W), V (F555W), and I (F814W), either with the WFC3/UVIS or, if already present in the archive, with ACS/WFC. The set of filters was dictated by three necessities: (1) separate bright stars from faint star clusters; (2) derive accurate ($\delta\tau\lesssim$10~Myr) SFHs from the CMDs; (3) obtain extinction--free ages and masses for clusters with age accuracy $\delta log(\tau)\sim$0.2 at intermediate ages, and mass accuracy $\delta log(M)\sim$0.3. The discrimination between faint clusters and massive stars will be performed via a combination of concentration index (concentration of the light within the central 1 pixel relative to 3 pixels radii) and U$-$B vs. V$-$I colors \citep[see Section~5.3 and][]{Chandar2010b}. 

The five LEGUS bands provide the minimum photometric set to break the age--dust extinction degeneracy in star clusters, and enable derivation of ages and masses  with the accuracy 
stated above, via SED--fitting on a cluster--by--cluster basis \citep[e.g.,][]{Adamo2012}. The U$-$B color is an effective age indicator, and the NUV filter replaces the more 
traditional H$\alpha$ filter 
as a discriminator between young and dusty clusters and old, dust--free clusters  \citep[e.g.,][]{Chandar2010b}. All clusters below $\sim$3,000--5,000~M$_{\odot}$ are subject to 
significant random (stochastic) sampling of the IMF, which first affects the ionizing photon rate of young star clusters \citep[e.g.,][]{Villaverde2010,Fouesneau2012,daSilva2012}. 
Like the ionizing photons, the NUV stellar continuum also traces massive stars, while providing more photometric stability (by a factor $\sim$3.5--4) relative to the H$\alpha$ emission \citep{Calzetti2010,Lee2011,Andrews2013}. This enables derivation of relatively accurate 
ages and masses of young ($\lesssim$10~Myr) star clusters down to $\sim$500--1,000~M$_{\odot}$, when the SED--fitting technique is combined with metallicity--matched, single--age stellar population models that include both deterministic and stochastic IMF sampling \citep{daSilva2012}.  We will be expanding the SED--fitting to include stellar rotation and binary evolution, as models become available \citep{Eldridge2009,Sana2012,Leitherer2014}.  
For star clusters more massive than $\sim$10$^4$~M$_{\odot}$, the break of the age--extinction degeneracy will be further aided by H$\alpha$ imaging, when available\footnote{At the time of writing, a Cycle--22 HST program, GO--13773 (P.I.: R. Chandar), has been approved, to observe a number of LEGUS galaxies in the WFC3/UVIS F657N filter (H$\alpha+$[NII]). If all observations are successful, a total of 34 LEGUS galaxies will have H$\alpha+$[NII] imaging available (46 pointings), between new WFC3 and archival ACS narrow-band images.}.

For the CMDs of individual stars, NUV and V bands are required for deriving SFHs optimized for the most recent 50--100~Myr. The F275W was chosen as the best compromise between maximizing detection of individual stars (which are sensitive to the absorption feature of the extinction curve at 0.2175~$\mu$m) and providing the longest NUV--U leverage for star clusters \citep[which are mostly insensitive to the  0.2175~$\mu$m bump because of dust geometry][]{Calzetti2000}.  

The observations were designed to reach a depth of m$_{AB}$(NUV)=26.0, with signal--to--noise$\sim$6, and comparable depths in the other filters, for the typical 
stellar crowding conditions discussed in section~3.  Translated into the more commonly used Vega magnitude scale, this limit corresponds to m$_{Vega}$(NUV)=24.50. 
Higher levels of crowding will generally limit the depth of the redder filters first. The goal is to detect: (1) 10$^4$~M$_{\odot}$, 100~Myr old clusters, with mean 
A$_V$(continuum)=0.25~mag, at a distance of 12~Mpc; or (2) 10$^4$~M$_{\odot}$, 1~Gyr old star clusters at 5~Mpc; (3) MS stars with minimum mass 7--10~M$_{\odot}$ (depending on distance and crowding conditions), with S/N=6 in the NUV; (4) SN Ib/c progenitors with $E(B-V) \approx 0.4$ in the 
NUV with S/N=5, out to 11~Mpc for binary progenitors and out to 6~Mpc for a  single early type WC, assuming $\log T_{eff}({\rm K})  \sim 4.6$.  
As shown in the next Sections, the required depth of m$_{AB}$(NUV)=26.0 was accomplished by our exposures, which were all taken with a minimum of 3 dither steps to both remove cosmic rays and fill in the gap in the WFC3/UVIS detectors. Table~\ref{tab3} lists, for each combination of primary and parallel filters, the typical exposure times, and the number of orbits employed.

\section{Data Processing and Products}

\subsection{Images and Mosaics}
The WFC3/UVIS datasets were processed through the {\sc calwf3} pipeline version 3.1.2 once all the relevant calibration files (bias and dark frames) for the date of observation were available in MAST. The calibrated, flat-fielded individual exposures (``FLT'' files) were corrected for charge transfer efficiency (CTE) losses by using a publicly available stand-alone program \footnote{Anderson, J., 2013, http://www.stsci.edu/hst/wfc3/tools/cte\_tools}. 
These corrections were small because we used the post-flash facility\footnote{http://www.stsci.edu/hst/wfc3/ins\_performance/CTE/ANDERSON\_UVIS\_POSTFLASH\_EFFICACY.pdf}  for the F275W, F336W, and F438W exposures to increase the background to a level near 12~e$^-$. At these levels, CTE losses represent a small perturbation on the charge transferred on readout of the CCDs. The resulting ``FLC'' files were then aligned and combined using the {\sc drizzlepac} software \citep{Gonzaga2012}. In summary, we first aligned and combined individual exposures for each filter, then aligned the combined images across filters using the F438W  (or F336W) image as the reference frame for the World Coordinate System (WCS). We then re-combined the exposures for each filter using the solutions determined from the alignments within and across filters to provide the final data products.

In detail for WFC3/UVIS data, the individual exposures for each filter were first aligned using the  {\sc tweakreg} routine. The shifts, scale and rotation of individual exposures were solved for using catalog matching to an accuracy of better then 0.1 pixels. Each catalog typically contains a few hundred sources that are used for 
determining the alignment solution. The {\sc astrodrizzle} routine was then used to combine the aligned images for each filter at the native pixel scale.
Each image was sky--subtracted\footnote{The sky-subtraction is automatically performed by {\sc astrodrizzle}, and consists of subtracting the mode from each image, before combination. Thus, for extended objects, the `sky--subtraction' step does not correspond to the removal of an actual sky value. The value of the mode (the subtracted `sky')  is stored in the image header, and can be readded to the data, if needed.}, weighted by its exposure time and a Gaussian kernel was used. The resulting cosmic--ray corrected, combined and drizzled images for each filter were then aligned with  {\sc tweakreg} to a common reference frame, using the WCS of the F438W image or the F336W image, if no F438W image was available. We chose to use a WFC3/UVIS image as the WCS reference because of the more accurate coordinates of Guide Star Catalog II used for more 
recent HST observations.
Next, a routine called {\sc tweakback} was used to propagate the new WCS solutions back to the aligned FLC images for each filter. The final step was to re-drizzle these images for each filter using sky subtraction, exposure time weighting, a Gaussian kernel and the UVIS native pixel scale. The final data products for the WFC3/UVIS data are in units of e$^-$\,s$^{-1}$ with a pixel scale of 39.62 mas\,pixel$^{-1}$ and are registered with North up and East to the left.

The units of e$^-$\,s$^{-1}$ enable a user to convert instrumental measurements to physical units using the WFC3 photometric zeropoints, which are included as keywords in the headers of the data products and are also posted at: http://www.stsci.edu/hst/wfc3/phot\_zp\_lbn. 

Mosaics were made as the final data product for those targets with multiple overlapping pointings (e.g. NGC\,4258, NGC\,5194$+$5195, etc.). The mosaicking was done with {\sc tweakreg} by aligning the drizzled images in the overlap region and propagating the solutions with {\sc tweakback}  to the FLC images before the final drizzle step.

For ACS/WFC data, the same data reduction procedure was followed. We retrieved the CTE--corrected data from MAST and aligned and drizzled the images using the ACS/WFC native pixel scale of 0.049 arcsec\,pixel$^{-1}$. The combined ACS/WFC images for each filter were then aligned to the UVIS WCS reference frame and re-drizzled to the native UVIS pixel scale to provide final data products that are equivalent to the UVIS data products. It was sometimes necessary to mosaic ACS images together to cover the UVIS field of view because of the different pointings of the archival images.

To execute the steps in the data reduction sequence described above, we developed automatic scripts to perform the image alignment. These are based on scripts created for the HST Frontier Fields project (D. Hammer, priv. comm.). The data reduction procedure that was adopted was extensively tested. We verified that data re--sampling through drizzling and rotating the images to N--E  has no affect on the photometric accuracy. Differences were well below 0.1 mag and thus smaller or comparable to (and, for faint sources, smaller than) typical photometric errors. We also compared aperture photometry (performed by subtracting the local background measured in a annulus close to the source) for sky--subtracted and unsubtracted datasets and found no differences above 0.1 mag.

\subsection{Stellar Photometry}

Stellar photometry is being performed on the individual, uncombined ``FLC'' frames, using the photometric package DOLPHOT 2.0 \citep{Dolphin2002}, with the WFC3 and/or ACS module. This package is designed to measure the flux of stars in dithered HST images acquired with the same position angle and small ($\la 30\arcsec$) shifts between exposures. The aligned FLC files contain the shifts derived by {\sc tweakreg} as header keywords; we use these as starting points
for DOLPHOT, which is then allowed to optimize the shifts among the images using bright stars that are common to all the images. 

The photometry is carried out independently in each filter. DOLPHOT iteratively identifies peaks and uses PSF models from Tiny Tim \citep{Krist1995, Hook2008} to simultaneously fit the PSF and the sky to every peak within a stack of images. Minor corrections for differences between the model and the real PSF in each exposure are calculated  using bright stars. DOLPHOT uses all the exposures in each filter to obtain stellar photometry, and for all the detected sources provides several parameters, including position, object type, average magnitude and magnitude error, signal-to-noise, sharpness, roundness, $\chi^2$ fit to the PSF, crowding, and error flags. 

DOLPHOT can apply an empirical CTE correction to the photometry. We decided to turn this option off, since our input images are already corrected for CTE losses (see previous Section). The most isolated stars in each filter are used to determine aperture corrections to the PSF magnitude, which account for differences between the model and the real image PSF. The final measured count rates are converted into the VEGAMAG system, by adding the zero points provided by the WFC3 team \footnote{ http://www.stsci.edu/hst/wfc3/phot\_zp\_lbn}.

At the end of this process, for each target we will obtain 5 catalogs, one for each band. We have tested several parameter combinations to remove as many spurious objects from our catalogs without affecting their completeness. In particular we select only the sources that are flagged by DOLPHOT as stars ($OBJTYPE=1$), 
have signal--to--noise ratios $\ge$3,  error\_flag$\le$1 (meaning the star is recovered without saturated pixels or other problems, see the DOLPHOT 2.0 Manual), 
and $\chi^2>1.2$. 
Band--merged cleaned catalogs are being produced by combining the single--band catalogs using the public cross-correlation algorithm CataPack\footnote{http://www.bo.astro.it/\textasciitilde{}paolo/Main/CataPack.html, developed at the INAF -- Bologna Observatory}.

A more complete description of the stellar photometry for the LEGUS galaxies, including artificial star tests to investigate completeness, blending, photometric errors, and effects of crowding, will be reported in \citet{Sabbi2014}.

\subsection{Cluster Photometry}

Due to the extended and often irregular nature of star clusters, a different approach from stellar photometry is being implemented for the identification 
and photometry of stellar clusters.

At the distance range of the LEGUS galaxies, clusters of typical half--light effective radii \citep[between 1 and 10 parsec,][]{PortegiesZwart2010} look like compact  
sources. However, they are more extended than stars: a compact cluster with r$_{eff}$=1~pc at a distance of 4~Mpc 
has a full-width half maximum (FWHM) of $\sim$2.5 WFC3/UVIS pixels, slightly broader than the stellar FWHM$\sim$2.2~pixels. Clusters in more distant 
galaxies will need to have larger effective radii to be discriminated from stars. Cluster detection and photometry has been 
optimized to detect resolved sources, as described below. When needed, we use as example the LEGUS test-bench galaxy NGC\,6503, which we also use to provide some preliminary results in the next Section.

For each galaxy, the photometric cluster catalogue is the result of a two--step process. The first step relies on a fully automatic approach. The source extractor algorithm SExtractor \citep{Bertin1996} is used with a parameter set optimized to select slightly extended sources on a variable background. In order to avoid the color biases in cluster selection that are produced by using a single filter (and that can result in age biases), source detection is being performed on white 
light images; these are the combination of the images in all 5 filters, weighted by the signal--to--noise (based on the median DOLPHOT photometric signal--to--noise).  The output from this first step contains not only candidate clusters but also bright single stars and background sources. To remove as many single bright stars as possible,  we perform a concentration index (CI) analysis, on the V band filter (WFC3/F555W or ACS/F555W). The CI is defined as the difference of the magnitude of each source at two different aperture radii, 1 px and 3 px, and it quantifies the concentration of the light in each object. In the NGC\,6503 data, the stellar CI has a typical narrow Gaussian distribution around the value 1.05 mag, while clusters have values larger than 1.25 mag. We expect the CI threshold for separating stars from more extended objects will be a function of the galaxy distance, and will change from galaxy to galaxy. For NGC\,6503, we have generated a catalogue of potential cluster candidates, which contains only sources with CI larger than 1.2 mag (slightly more conservative than the typical cluster CI of 1.25 mag). Using this catalogue, we have performed aperture photometry in all the 5 available bands, with a radius of 4 pixels (corresponding to 0.16$\arcsec$, or 4.1~pc at the distance of NGC\,6503, see Table~\ref{tab1}) and sky annulus with radius of 7 pixels, and 1 pixel wide. This automatic catalogue counts a total of $\sim$4600 objects, with a CI larger than 1.2 and detection in at least two contiguous filters with photometric error $\sigma \leq 0.3$ mag. The photometry is based on the Vega magnitude system. Galactic foreground extinction has been removed using the information available from the NASA Extragalactic Database (NED). Filter--dependent aperture corrections, in the range 0.7--0.8~mag, 
have been estimated using isolated clusters in each frame of NGC\,6503, and the published photometry already includes these corrections.

The second step has the aim of producing a  high--fidelity cluster catalog.  This step is a combination of (1) a semi-automatic approach which imposes 
additional, science--driven, selection criteria to the automatic catalog, in order to further reduce the number of spurious detections, and (2)  subsequent 
visual inspection of the individual candidates to provide confirmation of their nature through use of the multi-wavelength and high--angular--resolution 
information of the data. 

We include the following additional selection criteria in our automatic catalog: we 
require detection in at least 4 filters (each with error less than 0.3 mag) and (for NGC\,6503) V magnitude brighter than 22.6 mag, for all cluster candidates with CI$>$1.25. 
The first condition is imposed to obtain reliable constraints on the derived cluster properties (age, mass, extinction). Photometric information is needed in at least 4 different bands, 
with one point covering the spectrum of the cluster in the NUV or U, to be able to break the age--extinction degeneracy \citep[e.g.,][]{Anders2004,Chandar2010b,Konstantopoulos2013}. 
This selection criterion brings down the total number of clusters to be inspected from $\sim$4,600 to $\sim$3,000. The magnitude limit is introduced according to the  detection 
limits required by the LEGUS science goals: a 10$^4$~M$_{\odot}$, 100~Myr old cluster has an absolute luminosity of $\sim -7$ mag in the F555W filter.  We apply a 
brightness cut 1 magnitude fainter than this limit (i.e. $-6$~mag), which enables selecting down to $\sim$1,000~M$_{\odot}$, 6~Myr old clusters with color excess E(B--V)=0.25.  The 
magnitude cut is imposed on the {\em aperture corrected} F555W magnitudes. 
At the distance of NGC\,6503 this limit corresponds to a visual apparent magnitude of 22.6 mag. The apparent V magnitude limit will, obviously, vary with the galaxy distance, but will 
be maintained to an absolute V magnitude of $-6$~mag for all LEGUS galaxies. 
This second criterion reduces the number of candidate star clusters in NGC\,6503 from $\sim$3,000 to 402. 

Next, we use a custom--made, Python--based, visualization tool to inspect the candidate star clusters. In 
NGC\,6503, we have inspected all  402 cluster candidates, to which we have assigned one of four classes. The four classes are:  1, for a  centrally concentrated object; 2, for a concentrated object with some degree of asymmetry; 3, for a multiple peak system; 4, for a spurious detection (foreground/background sources, single bright stars, artifacts). These classes will be adopted for all cluster catalogs produced by LEGUS. Each cluster is visually inspected by 3 to 5 separate individuals, and we report in the catalog both the 
mode and the mean of the classes assigned by each individual.  We will consider star clusters in
classes 1, 2, and 3 as our `high fidelity' identifications. In NGC\,6503, classes 1, 2, and 3 include a total of 291 clusters (58, 92, 141 in classes 1, 2, 3, respectively, or 14\%, 23\%, and 
35\% of the total);  the remaining 111  objects (28\%) are in class 4 and are for the vast majority consistent with single unresolved sources, likely bright stars. 

The current (field--standard) approach of performing visual verification of each cluster limits the total number of clusters that can be inspected in 
each galaxy.  When extrapolating from NGC\,6503, our selection criteria yield an expected total of about 15,000--20,000 cluster candidates that will be visually inspected, across all 63 LEGUS pointings. This is already a significant number, and  larger numbers would require prohibitive effort\footnote{More automatic approaches are currently under investigation by the LEGUS team.}. However,  the full automatic, SExtractor--based catalogs will be released for all LEGUS galaxies, to enable more extended selection criteria to be applied. For the specific case of NGC\,6503, the SExtractor--based catalog contains approximately 4,600 sources, of which 
about 3,000 detected in the NUV or U and the three optical filters. For sources detected in at least four bands, physical information is added to the catalog: age, mass, extinction, and uncertainties, together with $\chi^2$ values, as derived from $\chi^2$-minimization--based SED--fitting (Section~6.2 for more details on the SED fits). Additional information is added in the catalog for the $\sim$400 visually inspected candidates, as detailed in Section~5.4. 

A detailed description of the cluster selection and identification procedure, including the SExtractor parameters set, and application to a variety of 
galactic morphologies and distances will be presented in \citet{Adamo2014}.  

\subsection{Data Products and Deliverables}

At the end of the project, a number of high--level products will be delivered to the community. The products will be initially hosted on a dedicated 
website (legus.stsci.edu) maintained by the LEGUS team; subsequently, the products will be migrated to stable archival platforms: MAST 
(http://archive.stsci.edu/) and the Hubble Legacy Archive (HLA, http://hla.stsci.edu/), or their successors. 

For each of the 50 LEGUS galaxies, the high--level data products will include:
\begin{enumerate}
\item Combined and aligned images in the 5 LEGUS bands (NUV,U,B,V,I), corrected for CTE losses and registered to a common WCS reference as given by the 
WFC3/UVIS/B (or U) band. When observations for a given galaxy include a mix of WFC3 and archival ACS images, the registration is performed relative to 
the WFC3 WCS. The final products have units of e$^-$\,s$^{-1}$ with a pixel scale of 39.62 mas\,pixel$^{-1}$ and are registered with North up and East to the left.
 \item Where available, narrow--band images in the light of the lines H$\alpha+$[NII] ($\lambda$6563~\AA\, $+$ 6548,6584\AA\AA) from archival ACS/WFC and/or 
 new WFC3/UVIS data processed in the same fashion as the broad--band images.
 \item When more than one overlapping pointing exists for a galaxy (or galaxy pair), a mosaic will be delivered, in addition to the processed individual pointings, with the same image characteristics as the individual pointings.
 \item Band--merged catalogs of unresolved sources detected in at least one band, flagged as stars by  DOLPHOT, and with  
 photometric errors, sharpness, roundness, and crowding within the limits described in Section~5.2. These catalogs, which we term `stellar catalogs', include:  
 positions in both X and Y and in RA(2000) and DEC(2000), aperture--corrected PSF--fitting photometry in 
 each of the five bands, together with their uncertainties, reduced $\chi^2$, sharpness, crowding, and roundness.  All these parameters are described in detail  
 in the  DOLPHOT documentation \citep{Dolphin2002}. Each source is also labeled with a LEGUS unique identifier.
\item Band--merged catalogs of resolved sources detected in at least two contiguous bands, as produced by SExtractor, and with a sufficiently large Concentration Index to 
exclude most stellar sources (the cut--off value of the CI index is distance--dependent and will change from galaxy to galaxy). These catalogs contain the most 
extensive selection of cluster candidates and include:  positions in both X and Y and RA(2000) and DEC(2000), aperture--corrected photometry in  each of the five bands, 
together with their uncertainties, and the source Concentration Index. Within the catalog,  clusters detected in at least four bands are identified via the Nflt flag (Nflt=4.0 or 5.0 for 
detection in four or five bands, respectively, each with photometric error $<$0.3 mag; remaining cluster candidates have Nflt=0.0). Visually--inspected clusters are identified via 
the value of the class (1, 2, 3, or 4) which has been attributed to them, as described in Section~5.3: ClMode for the class mode and ClMean for the class mean value (ClMode and 
ClMean are zero for  non--inspected clusters). For all clusters detected in at least four bands, both candidates and visually--inspected,  the best fitting age, mass, and color excess E(B--V), and 
their 68\% confidence levels are also listed as derived from $\chi^2$--minimization SED--fitting.  For each SED fit, a quality assessment is provided via: $\chi^2$ fit residuals for 
each band, reduced $\chi^2$ value for the all--bands fit, and a probability value Q (Q close to 1 implies a good fit, Q close to zero implies a poor or unconstrained fit). Although the 
parameters obtained from the fits are provided for all cluster candidates detected in at least four bands, only those parameters derived for the high--fidelity clusters, i.e., those 
with class 1, 2, or 3, should be considered reliable.
\item Existing ancillary imaging data for each galaxy. The minimum set of ancillary data includes: GALEX (two bands), SST (seven bands for 47 galaxies; five bands for 
NGC\,1433, NGC\,1566, and NGC\,6744), WISE (four bands), ground--based  R--band and continuum--subtracted H$\alpha$ (two bands, 41 galaxies). All of these data are already available from 
either MAST (e.g., GALEX) or IRSA (SST, WISE, and ground--based). However, the consolidation of the ancillary data will offer a one--stop--shop 
for the LEGUS galaxies\footnote{For WISE imaging, high--resolution mosaics, with a factor $\approx$3 improvement in the PSF relative to the native one, 
are being provided by T.H. Jarrett 
(private communication), following the technique described in \citet{Jarrett2012} and applied in \citet{Jarrett2013}. These mosaics will be provided as part of the LEGUS data products.}.
\end{enumerate}

\section{Initial Results}

\subsection{Stellar Populations}

The band--merged stellar catalog generated for the galaxy NGC\,6503 using the procedure described in Section~5.2 has been used to produce the 
Hess diagrams shown in Figures~\ref{fig4} and \ref{fig5}. 

Figure~\ref{fig4} illustrates how the data, in greyscale, compare with stellar synthetic models in a variety 
of color combinations, when the vertical (magnitude) scale is either the NUV (F275W) or the V (F555W). Although artificial stars tests have not been run 
yet to determine photometric errors, completeness, and blending, and the exact results may vary somewhat from the current representation, a few general 
features can be inferred. Despite the non--negligible level of crowding in this galaxy, the detection limit for the NUV filter is about m$_{Vega}(NUV)\sim$26.0, 
for sources detected with minimum S/N=3. This agrees with the survey observational goal of achieving m$_{Vega}(NUV)\sim$24.5 for sources with S/N=6 
(Section~4). Furthermore, in the blue filters combinations (e.g., NUV versus NUV$-$V) the bluest Blue--Loop excursions  remain to the red side 
of the Main Sequence, thus enabling a clean separation between the stars in the two evolutionary phases. As stated in Section~2.3, this is an important 
feature for deriving accurate recent--past SFHs in galaxies.

The optical (I versus V$-$I, Figure~\ref{fig5}) Hess diagrams for NGC\,6503 are compared with the Padova stellar evolutionary tracks 
\citep{Girardi2010}  at a range of metallicity values, from 
slightly--above solar\footnote{We adopt a solar metallicity Z=0.014 from \citet{Asplund2009}} (Z=0.017) down to about 1/35 solar. 
The location of the data relative to the tracks indicates that the stellar populations younger than $\approx$500~Myr are consistent with 
solar metallicity. 
A more detailed discussion will be presented in  \citet{Sabbi2014}. 

UV CMDs can be effectively employed to trace the clustering of young stellar populations. Figure~\ref{fig6} shows an example of the clustering of
the UV--bright, presumably young and massive, stars in NGC\,6503. The UV--bright population is identified as the region of the UV--U CMD delimited by
$-$2$\le$NUV$-$U$\le$2~mag and brighter than absolute magnitude M$_{NUV}=-$2.5~mag. The spatial distribution of these stars can be used to
compute surface density images, after smoothing to several scales, from $<$10~pc to $\sim$1~kpc. The smoothed surface density images are then subtracted from
each other to highlight localized over--densities at each scale; these are subsequently linked together in hierarchical structures relating spatially associated over--densities 
detected at any of the considered scales. The contours shown in Figure~\ref{fig6} represent the boundaries of significant over--densities defined using four selected smoothing 
kernels (scales) spanning more than an order of magnitude in size, stepping by a factor 2 difference in scale between contours. This
technique can be utilized to identify similarities and differences in the clustering of different stellar populations.  A more expanded version of this approach,
using star--by--star extinction--corrected CMDs and a range of galaxies, will be presented in \citet{Thilker2014}.

The angular two--point correlation function of the stars in NGC\,6503 shows a stronger correlation for stellar populations younger than $\sim$100~Myr than 
for stellar populations older than $\sim$500~Myr (Figure~\ref{fig7}). The older stars are almost homogeneously distributed across the galactic disk, 
while the young stars show a hierarchical pattern in their distribution with a correlation dimension $D_2 \sim 1.7$ \citep{Gouliermis2014c}.

\subsection{Star Cluster Populations}

Two examples of color--color diagrams (CCDs) for the clusters in NGC\,6503 are shown in Figure~\ref{fig8}, using UV and optical colors.  
The cluster candidates obtained from the automatic catalog (in the background) are compared with the visually--confirmed clusters in the high--fidelity sample (classes 1, 2, and 3, see Section~5.3). The colors of high--fidelity clusters have a significantly smaller scatter than those of the full automatic catalog, and are also closer to the expected colors 
of models of single--age stellar populations. The models cover the age range between 1~Myr and $>$1~Gyr, but the vast majority of the high--fidelity clusters are younger than 
a few 100~Myr, as per survey design. CCDs and corresponding CMDs (not shown here) are useful to obtain the ensemble picture of the distribution of the star clusters' ages, 
but the actual values of age, mass, and dust extinction are derived from the multi--band SED fitting of each cluster. 

Using the algorithm and error treatment described in \citet{Adamo2010,Adamo2012}, we show in Figure~\ref{fig9} the results of the SED fits of two class~1 star 
clusters in NGC\,6503: one relatively young ($\sim$6~Myr old) and one more evolved  ($\sim$100~Myr old). For these fits, we use models that implement  
deterministic sampling of the stellar IMF, since the masses are large enough ($>$10$^4$~M$_{\odot}$) that they are not much affected by stochastic IMF sampling.
The synthetic models used in the fits are 
those of \citet{Zackrisson2011},  that include both stellar and nebular emission. The latter component can have significant impact on young stellar populations, 
where both nebular lines and nebular continuum can be strong \citep[e.g.,][]{Reines2010}. 
We adopt solar metallicity models, guided by the results in the previous Section and by the fact that galaxies in the same morphological range as NGC\,6503 
tend to have solar or slightly below--solar metallicity.  The cluster masses are derived under the assumption of a \citet{Kroupa2001} IMF in the mass range 0.1--120~M$_{\odot}$. 
Conversion to a Salpeter IMF in the same mass range would require multiplying the cluster masses by a factor 1.6. In addition to the best--fit SEDs, Figure~\ref{fig9} contains 
the distribution of the $\chi^2$ values in the age--versus--E(B--V) and age--versus--mass planes. The E(B--V) values are derived using the attenuation curve of 
\citet{Calzetti2000}. For both clusters, we find a definite minimum value/region for the 
age--extinction--mass combination, demonstrating the power of the 5 LEGUS photometric bands in constraining these parameters in simple stellar populations. The 68\% 
confidence levels (red contours in Figure~\ref{fig9}) give uncertainties of less than 30\%, 20\%, and 10\%, respectively for the age, mass, and extinction of the  
intermediate (100~Myr) age cluster. The uncertainties are smaller for the younger (6~Myr) cluster, being at the level of $<$5\%, 10\%, and 25\% for the age, mass, 
and extinction, respectively. This level of accuracy is sufficient for most scientific applications. More details will be included in \citet{Adamo2014}. 

\section{Public Outreach}

An integral part of the LEGUS project is its outreach component, which is creating 3D tactile representations of galaxies. This 
approach is building on the experience gained by members of the LEGUS team on a previous, similar  project that uses star forming 
regions in the SMC. The main goal of the outreach component is to stimulate an understanding of astronomical phenomena in individuals 
who are visually impaired and/or are tactile learners, with a specific goal of reaching middle and high school students. The basic procedure 
is to  transform  multi--color Hubble images, like those obtained by LEGUS, into  3D models of astronomical objects, analogous to a visual fly-through, 
using 3D printers. This effort is being supported via a separate HST/EPO program (HST/EPO \# 123364, PI C. Christian).

\section{Summary}

LEGUS is an HST Cycle~21 Treasury program that is imaging 50 nearby galaxies in 5 broad--bands with the WFC3/UVIS, from the NUV to the I band.
The overall science goal is to link star formation across all scales, from individual stars to the multi--kpc scales of whole galaxies, through the full 
range of structures that newly formed stars occupy. The `tools' to achieve this goal include, but are not limited to: the investigation of the hierarchical star formation, 
including dissipation timescales; the evolution and disruption of star clusters; the recent star formation histories of galaxies. LEGUS will also provide a 
census of the UV--bright globular clusters across a range of environments, and a reference database for future identification and study of the progenitors of 
supernovae. 

The science results from LEGUS will inform models and investigations of the evolution of the luminous baryonic component of galaxies across cosmic 
times. To this end, we will be releasing to the community a number of higher--level products, including multi--color images, mosaics, and photometric 
catalogs for both stellar sources and star clusters. For the clusters, we will also release catalogs of physical properties, including ages, masses, and 
extinction values. These latter catalogs will be unprecedented, as no such lists of physical characteristics of cluster populations for a large number of 
galaxies currently exist in the public domain. The LEGUS observations and data products will provide a foundation for future investigations of nearby 
and distant galaxies and star formation with ALMA and the JWST.

\acknowledgments

Based on observations made with the NASA/ESA Hubble Space Telescope, obtained  at the Space Telescope Science Institute, which is operated by the 
Association of Universities for Research in Astronomy, Inc., under NASA contract NAS 5--26555. These observations are associated with program \# 13364. 
Support for program \# 13364 was provided by NASA through a grant from the Space Telescope Science Institute.

This research has made use of the NASA/IPAC Extragalactic Database (NED) which is operated by the Jet
Propulsion Laboratory, California Institute of Technology, under contract with the National Aeronautics and Space
Administration.

S.d.M. acknowledges support for this work by NASA through an Einstein 
Fellowship grant, PF3-140105. C.L.D. acknowledges funding from the European Research Council for the
FP7 ERC starting grant project LOCALSTAR. D.A.G. kindly acknowledges financial support by the German Research 
Foundation through grant GO\,1659/3-1. A.H. acknowledges support by the Spanish MINECO under project
grant AYA2012-39364-C02-1. J.E.R. gratefully acknowledges the support of the National Space 
Grant College and Fellowship Program and the Wisconsin Space Grant 
Consortium. A.W. acknowledges funding from the European Research Council under 
the European Community's Seventh Framework Programme (FP7/2007--2013 
Grant Agreement no. 321323).




\clearpage

\begin{deluxetable}{lrrrrrrllrrrrrr}
\tablecolumns{15}
\rotate
\tabletypesize{\scriptsize}
\tablecaption{Properties of the LEGUS Galaxies Sample.\label{tab1}}
\tablewidth{0pt}
\tablehead{
\colhead{Name\tablenotemark{a}} & \colhead{v$_H$\tablenotemark{a}} 
& \colhead{Morph.\tablenotemark{a}}  & \colhead{T\tablenotemark{b}} & \colhead{Inclin.\tablenotemark{a}} 
& \colhead{Dist.\tablenotemark{c}} & \colhead{Method\tablenotemark{d}}
& \colhead{Ref\tablenotemark{e}} & \multicolumn{2}{c}{12$+$log(O/H)\tablenotemark{f}} 
& \colhead{Ref\tablenotemark{g}}
& \colhead{SFR(UV)\tablenotemark{h}} & \colhead{M$_*$\tablenotemark{i}} 
& \colhead{M(HI)\tablenotemark{j}} & \colhead{Ref\tablenotemark{k}}
\\
\colhead{} & \colhead{(km~s$^{-1}$)} 
& \colhead{} & \colhead{} & \colhead{(degrees)} 
& \colhead{(Mpc)} & \colhead{} 
& \colhead{}  & \colhead{(PT)} & \colhead{(KK)} & \colhead{} 
& \colhead{(M$_{\odot}$~yr$^{-1}$)} & \colhead{(M$_{\odot}$)}
& \colhead{(M$_{\odot}$)} & \colhead{}  
\\
\colhead{(1)} & \colhead{(2)} & \colhead{(3)} & \colhead{(4)} & \colhead{(5)} & 
\colhead{(6)} & \colhead{(7)} & \colhead{(8)} & \colhead{(9)} & \colhead{(10)} &
\colhead{(11)} & \colhead{(12)} & \colhead{(13)} & \colhead{(14)} & \colhead{(15)}
\\
}
\startdata
\multicolumn{15}{c}{T=0--2\tablenotemark{l}}\\
\hline
NGC1291 & 839 & SBa           & 0.1(0.4) &  34.3 & 10.4  & TF   &  1  & 8.52(+) & 9.20(+)   & 1 & 0.63 &  1.5E11  & 2.3E09 &  1 \\ 
NGC1433 & 1076 & SBab      & 1.5(0.7) & 24.8 &  8.3  & TF   & 2   & \nodata  & \nodata  &  & 0.27 &  1.7E10  & 5.0E08 & 1 \\
NGC1510 & 913 & SA0\tablenotemark{m} &$-$1.6(1.7)\tablenotemark{m}& 0.0 & 11.7 & TF & 3   &  \nodata &   8.38   & 4 & 0.12 &  4.8E08  & 6.5E07 & 2 \\
NGC1512 & 896& SBab          & 1.1(0.5) &  51.0 & 11.6  & TF   &  3  & 8.56    & 9.11          & 1 & 1.00 &  1.7E10  &  8.5E09 & 2  \\ 
ESO486--G021 & 835 & S?    & 2.0(1.7) & 48.2 & 9.5 & v(flow) & \nodata & \nodata& \nodata  & &  0.05 & 7.2E08  & 2.8E08 & 1 \\
NGC3368 & 897 & SABab      & 1.9(0.6) & 46.8 & 10.50 & Ceph & 4  &  \nodata &  9.04  & 3  &  1.10 & 4.8E10  & 2.7E09 & 3 \\
NGC4594 & 1024 & SAa         & 1.1(0.3) &  66.3 &  9.1 & SBF  & 5  & 8.54(+) & 9.22(+)  & 1 &  0.48 & 1.5E11  &  2.8E08 & 4  \\ 
\hline
\multicolumn{15}{c}{T=2--4}\\
\hline
NGC1566 & 1504 & SABbc   &4.0(0.2) & 37.3 & 13.2 & TF & 6  & 8.63(+)  &   9.64(+) & 1 &  5.67  & 2.7E10 & 5.7E09 & 1\\
NGC3351 & 778 & SBb          &3.1(0.4) & 21.3 &  10.00 & Ceph &  4 & 8.60    & 9.19 & 1 & 1.57  & 2.1E10 & 1.3E09  & 3 \\ 
NGC3627 & 727 & SABb       &3.1(0.4) &  62.5 &  10.10 & Ceph &  4  & 8.34    & 8.99& 1 & 4.89  &  3.1E10 &  1.5E09  & 3  \\ 
NGC4248 & 484 & S?             &3.3(2.9) & 71.2 & 7.8 & TF & 7              & \nodata & 8.15 & 7  & 0.02  & 9.8E08  &  6.1E07 & 5  \\   
NGC4258 & 448 & SABbc      &4.0(0.2) & 67.2  & 7.98 & Ceph & 4      & \nodata &  8.89& 3 & 2.51  & 2.9E10 & 7.3E09 & 4 \\
IC4247     &  274 &  S?            &2.2(3.5) &  67.4 & 5.11 & TRGB & 8     & \nodata &  8.27   & 8 & 0.008 & 1.2E08 & \nodata & \nodata \\
NGC5195 & 465 & SBa          &2.2(4.5) &37.5 & 7.66 & SBF & 9       &  8.36(+) & 8.99(+) & 1  & 0.35  & 2.3E10 & 1.7E09 & 3 \\
\hline
\multicolumn{15}{c}{T=4--6}\\
\hline
UGC0695 & 628 & Sc            &6.0(2.0) & 35.1   & 10.9 & v(flow) & \nodata & \multicolumn{2}{c}{7.69}   & 5  & 0.02 & 1.8E08  & 1.1E08 & 3 \\
NGC0628 & 657 & SAc         &5.2(0.5) & 25.2  &  9.9  & SNII &  10 & 8.35    & 9.02            & 1 & 3.67  & 1.1E10 &  1.1E10 & 4  \\ 
IC0559      & 514 & Sc            &5.0(3.0) &  41.4  & 5.3 & v(flow) & \nodata      &  \multicolumn{2}{c}{8.07}  & 5 & 0.005 &1.4E08 &  3.7E07 & 3 \\
NGC3344  & 580 & SABbc   &4.0(0.3) & 23.7 & 7.0 & v(flow) & \nodata    & 8.43 & 8.76  & 15, 9 & 0.86  & 5.0E09 & 2.3E09 & 4 \\
NGC4605  & 136 &SBc         &5.1(0.7) & 67.7  & 5.70 & TRGB & 8               & \nodata  & 8.77  & 9 & 0.43  & 1.5E09 & 3.7E08 & 4 \\
NGC5194  & 463 & SAbc      &4.0(0.3) & 51.9 & 7.66 & SBF & 9                &  8.55  &  9.18  & 1 & 6.88  & 2.4E10 & 2.3E09 & 6 \\
NGC5457 & 241 & SABcd    &6.0(0.3) & 20.9 &  6.70  & Ceph &  4 & \multicolumn{2}{c}{8.48} & 2  & 6.72  &1.9E10 & 1.9E10 & 7  \\ 
NGC5949 & 430 & SAbc       &4.1(0.3) & 65.5  & 14.3 & TF & 2          &  \nodata & \nodata  & & 0.38  & 1.8E09 & 2.8E08 & 3\\
NGC6503 &   25  & SAcd      &5.8(0.5) & 70.2  & 5.27 & TRGB & 11  &  \nodata   &   \nodata  & & 0.32   & 1.9E09 & 1.3E09 & 8\\
NGC6744 & 841 & SABbc    &4.0(0.2) & 44.1 & 7.1 & TF & 2            &   8.55   &     \nodata    & 10 & 6.48  & 2.2E10 & 1.2E10 & 1\\
\hline
\multicolumn{15}{c}{T=6--8}\\
\hline
NGC0045 & 467 & SAdm     &7.8(0.7) & 46.0 & 6.61 & TRGB & 8     &   \nodata  &  \nodata   & & 0.35  & 3.3E09 & 2.5E09 & 3 \\
NGC1313 & 470 & SBd        &7.0(0.4) & 40.7 & 4.39 & TRGB & 8        &   \multicolumn{2}{c}{8.4}  & 11 & 1.15  & 2.6E09  & 2.1E09 & 1 \\
NGC2500 & 504 & SBd        &7.0(0.3) & 26.3 & 10.1 & TF & 7              &  \nodata  &   8.84    & 9 & 0.46  & 1.9E09 &  8.2E08 & 3 \\
NGC3274 & 537 & SABd      &6.6(0.6) & 61.6 & 6.55 & BS & 12        &   \multicolumn{2}{c}{8.33}   & 12 & 0.07  & 1.1E08 & 5.5E08 & 4 \\
UGC7242 &   68  & Scd         &6.4(1.3) & 65.1 & 5.42 & TRGB & 8       &  \nodata  &  \nodata   & & 0.007 & 7.8E07 & 5.0E07 & 9 \\
NGC4242 & 506 & SABdm   &7.9(0.5) & 40.5 & 5.8 & TF & 2          &  \nodata  &  \nodata    & & 0.10   & 1.1E09  & 3.5E08  & 3 \\
NGC4490 & 565 & SBd         &7.0(0.2) & 60.5 & 7.2 & TF & 6              &  \multicolumn{2}{c}{8.35}   & 14 & 1.99  &  1.9E09 & 2.7E09 & 10 \\
NGC5238 & 235 & SABdm   &8.0(0.5) & 53.9 & 4.51 & TRGB & 8  &    \nodata         &  8.66   & 9  & 0.01  & 1.4E08 & 2.9E07 & 4 \\
NGC5474 & 273 & SAcd       &6.1(0.5) & 26.4 &  6.8  & BS   & 13      & 8.31      & 8.83        & 1 &  0.27 & 8.1E08 & 1.3E09 & 4 \\ 
NGC7793 & 230 & SAd         &7.4(0.6) & 47.4 &  3.44 & Ceph & 14   & 8.31     & 8.88        & 1 & 0.52 & 3.2E09  & 7.8E08 &  1 \\ 
\hline
\multicolumn{15}{c}{T=8--9.5}\\
\hline
UGC0685 & 157 & SAm       &9.2(0.8) & 41.4 & 4.83 & TRGB & 8        &  \multicolumn{2}{c}{8.00}& 6  & 0.007 & 9.5E07  & 9.7E07 & 3 \\
UGC1249 & 345 & SBm       &8.9(0.6) & 63.3 & 6.9 & TF & 15               & \nodata  &  8.73  & 9 & 0.15   & 5.5E08 & 9.9E08 & 11\\
UGC7408 & 462 & IAm         &9.3(2.8) & 62.5 & 6.7 & TF & 7                  &  \nodata  &  \nodata   & &  0.01 & 4.7E07 & 8.6E07 & 12\\
NGC4395 & 319 & SAm       &8.9(0.4) & 33.6 & 4.30 & Ceph & 16        &  8.26   &    \nodata     & 15 & 0.34 & 6.0E08 & 1.8E09 & 3 \\
NGC4485 & 493 & IBm         &9.5(1.3) & 45.9 & 7.6 & v(flow) & \nodata &  \nodata  &  \nodata  & & 0.25 & 3.7E08 & 4.0E08 & 10 \\
NGC4656 & 646 & SBm       &9.0(0.7) & 0.  & 5.5 & TF & 2                      &  \multicolumn{2}{c}{8.09}    & 5 & 0.50  & 4.0E08 & 2.2E09 & 4 \\
NGC5477 & 304 & SAm       &8.8(0.5) & 40.1 & 6.4 & TF & 7                  &   \multicolumn{2}{c}{7.95}     & 5 & 0.03  & 4.0E07 & 1.3E08 & 4\\
\hline
\multicolumn{15}{c}{T=9.5--11}\\
\hline
NGC1705 & 633 & SA0/BCG &11(...)& 42.5 & 5.1 & TRGB & 17           &  7.96     &   8.28     & 1 & 0.11  & 1.3E08 & 9.4E07 & 1 \\
UGC4305 & 142 & Im              &9.9(0.5) &  37.1 &  3.05 & Ceph & 18      &   \multicolumn{2}{c}{7.92} & 13 & 0.12  & 2.3E08  &  7.3E08 & 4  \\  
UGC4459 & 20   & Im              &9.9(0.5) &  29.9 &  3.66 & TRGB & 8        &  \multicolumn{2}{c}{7.82} & 13 & 0.007 & 6.8E06 & 6.8E07 &  9 \\  
UGC5139 &139 & IABm         &9.9(0.3) &  33.6 &  3.98  & TRGB & 8    &  \multicolumn{2}{c}{8.00}    & 13 &  0.02  & 2.5E07 &  2.1E08 &  3 \\  
UGC5340 & 503 & Im             &9.7(1.0) & 68.3 & 5.9 & TF & 7                   &  \multicolumn{2}{c}{7.20} & 5 & 0.02  & 1.0E07 & 2.4E08 & 4 \\ 
NGC3738 & 229 & Im             &9.8(0.7) & 40.5  & 4.90 & TRGB & 19        &   \multicolumn{2}{c}{8.04}  & 5 & 0.07  &  2.4E08 & 1.5E08 & 4\\
UGCA281 & 281 & Sm           &10.0(2.0)& 41.1 & 5.90 & TRGB & 20          &  \multicolumn{2}{c}{7.82}  & 16  & 0.02  & 1.9E07 & 8.3E07 & 4\\
NGC4449 & 207 & IBm          &9.8(0.5) & 44.8 & 4.31 & TRGB & 8           &   \multicolumn{2}{c}{8.26} & 5 & 0.94  & 1.1E09 & 2.1E09 & 4\\
NGC5253 & 407 & Im             &11(...) & 67.7 & 3.15 & Ceph & 4               &  \multicolumn{2}{c}{8.25}  & 17 & 0.10  & 2.2E08 & 1.0E08 & 1\\
\enddata

\tablenotetext{a}{Galaxy name, recession velocity, and morphological type as listed  in NED, the NASA Extragalactic Database. Inclination, in degrees,  derived from the sizes listed in NED.}
\tablenotetext{b}{RC3 morphological T--type as listed in Hyperleda (http://leda.univ-lyon1.fr), and discussed in \cite{Kennicutt2008} for the LVL galaxies, from which the LEGUS sample is derived. In that paper, T--type =11 is adopted for galaxies mis--classified as early types, while being compact  irregular or Blue Compact Galaxies (BCGs). 
Uncertainties  on the morphological classification are in parenthesis. Some of the galaxies have large uncertainties, and they may be mis--classified.}
\tablenotetext{c}{Redshift--independent distance in Mpc, or flow--corrected redshift--dependent distance (v(flow) in Mpc, adopting H$_o$=70~km~s$^{-1}$~Mpc$^{-1}$.}
\tablenotetext{d}{Methods employed to determine the distances. In order of decreasing 
preference: Cepheids (Ceph), Tip of the Red Giant Branch Stars (TRGB), Surface Brightness Fluctuations 
(SBF), Supernova Type II Plateau (SNII), flow--corrected Tully--Fisher relation (TF),  and brightest stars (BS). For the flow--corrected, redshift-dependent distances the flow model of \citet{Karachentsev1996} is adopted, as described in Kennicutt et al. 2008.}
\tablenotetext{e}{References to the distances: 
1   -- \citet{Masters2005}; 
2   -- \citet{Tully2009}; 
3   -- \citet{Springbob2009};  
4   -- \citet{Freedman2001}; 
5   -- \citet{Jensen2003}; 
6   -- \citet{Theureau2007}; 
7   -- \citet{Tully1988}; 
8   -- \citet{Jacobs2009};
9   -- \citet{Tonry2001}; 
10  -- \citet{Olivares2010};
11 -- \citet{Karachentsev2003}; 
12  -- \citet{Makarova1998};
13  -- \citet{Drozdovsky2000};
14  -- \citet{Pietrzynski2010};
15  -- \citet{Nasonova2011}; 
16  -- \citet{Thim2004};
17  -- \citet{Tosi2001};
18  -- \citet{Hoessel1998}; 
19  -- \citet{Karachentsev2003b}; 
20  -- \citet{Schulte-Ladbeck2001}.}
\tablenotetext{f}{Characteristic oxygen abundances of the galaxies. For spirals, this is the globally--averaged abundance \citep{Moustakas2010}. The two columns, 
     (PT) and (KK), are the oxygen abundances on two calibration scales:  the  PT value, in the left--hand--side column, is from the empirical calibration of 
     \citet{PilyuginThuan2005}; the KK value, in the right--hand--side column, is from  the theoretical calibration of \citet{Kobulnicky2004}. When only one oxygen abundance 
     is available, and its attribution is uncertain or it is derived from the `direct' method \citep[i.e., T$_e$(OIII)--based abundances,][]{Kennicutt2003, Thuan2005, PilyuginThuan2007,  Croxall2009, Berg2012,MonrealIbero2012}, the value straddles the two columns.}
\tablenotetext{g}{References to the oxygen abundances:
1 -- \citet{Moustakas2010}, their Table~9 -- a (+) indicates oxygen abundance from the Luminosity-Metallicity relation;  
2 -- \citet{Kennicutt2003}; 
3 -- \citet{Bresolin1999};
4 -- \citet{Storchi-Bergmann1994}; 
5 -- \citet{Berg2012};
6 -- \citet{vanZee2006}; 
7 -- \citet{Kewley2005};  
8 -- \citet{Lee2007};
9 -- Using equation~18 in \citet{Kobulnicky2004} on the line fluxes in \citet{Moustakas2006};
10 -- \citet{Pilyugin2006};
11 -- \citet{Walsh1997}; 
12 -- \citet{Hunter1999}; 
13 -- \citet{Croxall2009}; 
14 -- \citet{PilyuginThuan2007};
15 -- \citet{Pilyugin2004}; 
16 -- \citet{Thuan2005};
17 -- \citet{MonrealIbero2012}.}
\tablenotetext{h}{Star Formation Rate (M$_{\odot}$~yr$^{-1}$), calculated from the GALEX 
     far--UV, corrected for dust attenuation as described in \citet{Lee2009}.}
\tablenotetext{i}{Stellar masses (M$_{\odot}$) obtained from the extinction--corrected B--band luminosity, and color information, 
using the method described in \citet{Bothwell2009}, and based on the mass--to--light ratio models of \citet{Bell2001}.}
\tablenotetext{j}{HI masses, using the line fluxes listed in NED, applying the standard formula: M(HI)[M$_{\odot}$] = 2.356$\times$10$^5$ D$^2$ S, where D is the distance in Mpc, and S if the integrated 21--cm line flux in units of Jy~cm~s$^{-1}$.}
 \tablenotetext{k}{References for the HI line fluxes, as follows:  
 1 -- \citet{Koribalski2004}; 
 2 -- \citet{Koribalski2009};
 3 -- \citet{Springbob2005}, using their self-absorption corrected values, when available;
 4 -- \citet{Huchtmeier1989};
 5 -- \citet{deVaucolulers1991};
 6 -- \citet{Walter2008};
 7 -- \citet{Paturel2003};
 8 -- \citet{Greisen2009};
 9 -- \citet{Begum2008};
 10 -- \citet{Kovac2009};
 11 -- \citet{Saintonge2008};
 12 -- \citet{Borthakur2011}.}
 \tablenotetext{l}{The galaxies are grouped according to their RC3 morphological T--type. Within each 
group, the galaxies are listed in order of increasing right ascension.}
\tablenotetext{m}{The morphological types of NGC1510 do not necessarily capture the true nature of this galaxy which has a high level of star formation in its center 
\citep[e.g.,][]{Meurer2006}.}
 \end{deluxetable}

\clearpage

\begin{deluxetable}{llrll}
\tablecolumns{5}
\rotate
\tabletypesize{\footnotesize}
\tablecaption{Observations.\label{tab2}}
\tablewidth{0pt}
\tablehead{
\colhead{Name} & \colhead{WFC3 (Primary)\tablenotemark{a}} 
& \colhead{\# Pointings\tablenotemark{a}} 
& \colhead{ACS (Parallel)\tablenotemark{b}} 
& \colhead{ACS (Archival)\tablenotemark{c}} 
\\
\colhead{(1)} & \colhead{(2)} & \colhead{(3)} & \colhead{(4)} & \colhead{(5)}  
\\
}
\startdata
\multicolumn{5}{c}{T=0--2}\\
\hline
NGC1291 & F275W,F336W & 1 & F435W,F606W,F814W & F435W,F555W,F814W\\
NGC1433 & F275W,F336W,F438W,F555W,F814W & 1 & F435W,F814W & \nodata \\
NGC1510\tablenotemark{d} &F275W,F336W,F438W,F555W,F814W & 1 & F435W,F814W & \nodata \\
NGC1512\tablenotemark{d} &F275W,F336W,F438W,F555W,F814W & 2 & F435W,F814W & \nodata \\
ESO486-G021&F275W,F336W,F438W,F555W,F814W & 1 & F435W,F814W & \nodata \\
NGC3368 & F275W,F336W,F438W,F555W,F814W & 1 & F435W,F814W & \nodata \\
NGC4594 & F275W,F336W,F814W & 1 & F435W,F814W & F435W,F555W\\
\hline
\multicolumn{5}{c}{T=2--4}\\
\hline
NGC1566 & F275W,F336W,F438W,F555W,F814W & 1 & F435W,F814W & \nodata \\
NGC3351 & F275W,F336W,F438W,F555W,F814W & 1 & F435W,F814W & \nodata \\
NGC3627 & F275W,F336W,F438W,F555W,F814W & 1 & F435W,F814W & \nodata \\
NGC4248 & F275W,F336W,F438W,F555W,F814W & 1 & F435W,F814W & \nodata \\
NGC4258\tablenotemark{e}  & F275W,F336W,F438W,(F555W,F814W) & 2 & F435W,F814W & F555W,F814W\\
IC4247 & F275W,F336W,F438W & 1 & -- & F606W,F814W\\
NGC5195\tablenotemark{f} & F275W,F336W & 1 & F435W,F606W,F814W & F435W,F555W,F814W\\
\hline
\multicolumn{5}{c}{T=4--6}\\
\hline
UGC0695 & F275W,F336W,F438W,F555W,F814W & 1 & F435W,F814W & \nodata \\
NGC0628\tablenotemark{g}  & F275W,F336W,(F555W) & 2 & F435W,(F606W),F814W & F435W,(F555W),F814W\\
IC0559     & F275W,F336W,F438W,F555W,F814W & 1 & F435W,F814W & \nodata \\
NGC3344 & F275W,F336W,F438W,F555W,F814W & 1 & F435W,F814W & \nodata \\
NGC4605 & F275W,F336W,F438W,F555W,F814W & 1 & F435W,F814W & \nodata \\
NGC5194\tablenotemark{f} & F275W,F336W & 3 & F435W,F606W,F814W & F435W,F555W,F814W\\
NGC5457 & F275W,F336W & 5 & F435W,F606W,F814W & F435W,F555W,F814W\\
NGC5949 & F275W,F336W,F438W,F555W,F814W & 1 & F435W,F814W & \nodata \\
NGC6503 & F275W,F336W,F438W,F555W,F814W & 1 & F435W,F814W & \nodata \\
NGC6744 & F275W,F336W,F438W,F555W,F814W & 2 & F435W,F814W & \nodata \\
\hline
\multicolumn{5}{c}{T=6--8}\\
\hline
NGC0045 & F275W,F336W,F438W,F555W,F814W & 1 & F435W,F814W & \nodata \\
NGC1313 & F275W,F336W & 2 & F435W,F606W,F814W & F435W,F555W,F814W\\
NGC2500 & F275W,F336W,F438W,F555W,F814W & 1 & F435W,F814W & \nodata \\
NGC3274 & F275W,F336W,F438W,F555W,F814W & 1 & F435W,F814W & \nodata \\
UGC7242 & F275W,F336W,F438W & 1 & F435W,F814W & F606W,F814W\\
NGC4242 & F275W,F336W,F438W,F555W,F814W & 1 & F435W,F814W & \nodata \\
NGC4490 & F275W,F336W,F438W,F555W,F814W & 1 & F435W,F814W & \nodata \\
NGC5238 & F275W,F336W,F438W & 1 & F435W,F814W & F606W,F814W\\
NGC5474 & F275W,F336W,F438W & 1 & F435W,F814W & F606W,F814W\\ 
NGC7793 & F275W,F336W,F438W,(F555W,F814W) & 2 & F435W,F814W & F555W,F814W\\
\hline
\multicolumn{5}{c}{T=8--9.5}\\
\hline
UGC0685 & F275W,F336W,F438W & 1 & F435W,F814W & F606W,F814W\\ 
UGC1249 & F275W,F336W,F438W & 1 & F435W,F814W & F606W,F814W\\ 
UGC7408 & F275W,F336W,F438W & 1 & F435W,F814W & F606W,F814W\\ 
NGC4395 & F275W,F336W,F438W,(F555W,F814W) & 2 & F435W,F814W & F555W,F814W\\
NGC4485 & F275W,F336W,F814W & 1 & F435W,F814W & F435W,F606W\\
NGC4656 & F275W,F336W,F438W,F555W,F814W & 1 & F435W,F814W & \nodata \\
NGC5477 & F275W,F336W,F438W & 1 & F435W,F814W & F606W,F814W\\ 
\hline
\multicolumn{5}{c}{T=9.5--11}\\
\hline
NGC1705 & F275W,F336W,F438W,F555W,F814W & 1 & F435W,F814W & \nodata \\
UGC4305 & F275W,F336W,F438W & 1 & F435W,F814W & F555W,F814W\\ 
UGC4459 & F275W,F336W,F438W & 1 & F435W,F814W & F555W,F814W\\ 
UGC5139 & F275W,F336W,F438W & 1 & F435W,F814W & F555W,F814W\\ 
UGC5340 & F275W,F336W,F438W & 1 & F435W,F814W & F606W,F814W\\ 
NGC3738 & F275W,F336W,F438W & 1 & F435W,F814W & F606W,F814W\\ 
UGCA281 & F275W,F336W,F438W & 1 & F435W,F814W & F606W,F814W\\ 
NGC4449 & F275W,F336W & 1 & F435W,F606W,F814W & F435W,F555W,F814W\\
NGC5253 & F275W,F336W & 1 & F435W,F606W,F814W & F435W,F555W,F814W\\
\enddata

\tablenotetext{a}{The filters used for the primary LEGUS WFC3/UVIS observations of each galaxy, and the number of pointings in the galaxy.}
\tablenotetext{b}{The filters used for the parallel ACS/WFC observations.}
\tablenotetext{c}{The filter for the available observations from the MAST archive; these are usually ACS/WFC.}
\tablenotetext{d}{The pointing of NGC1510 and the two pointings of NGC1512 were joined into a single strip starting from the center of NGC1512 and ending at the center of NGC1510.}
\tablenotetext{e}{One of the two pointings of NGC4258, of NGC7793, and of NGC4395 has been observed in three filters (with the remaining two filters available from the archive), while the other pointing has been observed in all five filters.}
\tablenotetext{f}{The pointing of NGC5195 was joined to those of NGC5194 in a single mosaic. The shape of the LEGUS pointings for these two galaxies reflect the existence of planned GO observations (GO--13340) that will cover the 
nucleus of NGC5194 with identical filters for the primary exposures.}
\tablenotetext{g}{Each of the two pointings of NGC0628 has been observed in three(two) filters, with exposures for the remaining two(there) filters available from the archive.}
\end{deluxetable}

\clearpage

\begin{deluxetable}{llllr}
\tablecolumns{5}
\rotate
\tabletypesize{\small}
\tablecaption{Exposure Times and Orbits.\tablenotemark{a}\label{tab3}}
\tablewidth{0pt}
\tablehead{
\colhead{ WFC3 (Primary)} 
& \colhead{Exposure Times} 
& \colhead{ACS (Parallel)} 
& \colhead{Exposure Times}  & \colhead{\# of Orbits} 
\\
\colhead{Filters} & \colhead{(s)} & \colhead{Filters} & \colhead{(s)}  & \colhead{} 
\\
}
\startdata
F275W,F336W,F438W,F555W,F814W & 2400,1100,900,1100,900 & F435W,F814W & 1400,620& 3 \\
F275W,F336W,F438W & 2400,1100,900 & F435W,F814W & 1400,520 & 2\\ 
F275W,F336W & 2500,2400 & F435W,F606W,F814W & 1500,1100,1400 & 2\\
\enddata

\tablenotetext{a}{Number of orbits and typical exposure times for each combination of filters for the targets listed in Table~\ref{tab2}, and for both primary and parallel observations. 
The filters listed in the second row are examples; while the NUV and U filters have been obtained for all galaxies, the third filter had been chosen to complement what present in the MAST archive. Each exposure time in the primary observations is split into 3 dithered sub--exposures. The parallel exposures are generally obtained with 2 dithered sub--exposures, except the 5--filters I--band (F814W) case, which has 3 sub--exposures. The sum of the primary exposures is typically longer than 
the sum of the parallel exposures, as priority was given to maximizing the former.}
\end{deluxetable}

\clearpage

\clearpage 

\clearpage 
\begin{figure}
\figurenum{3a}
\plotone{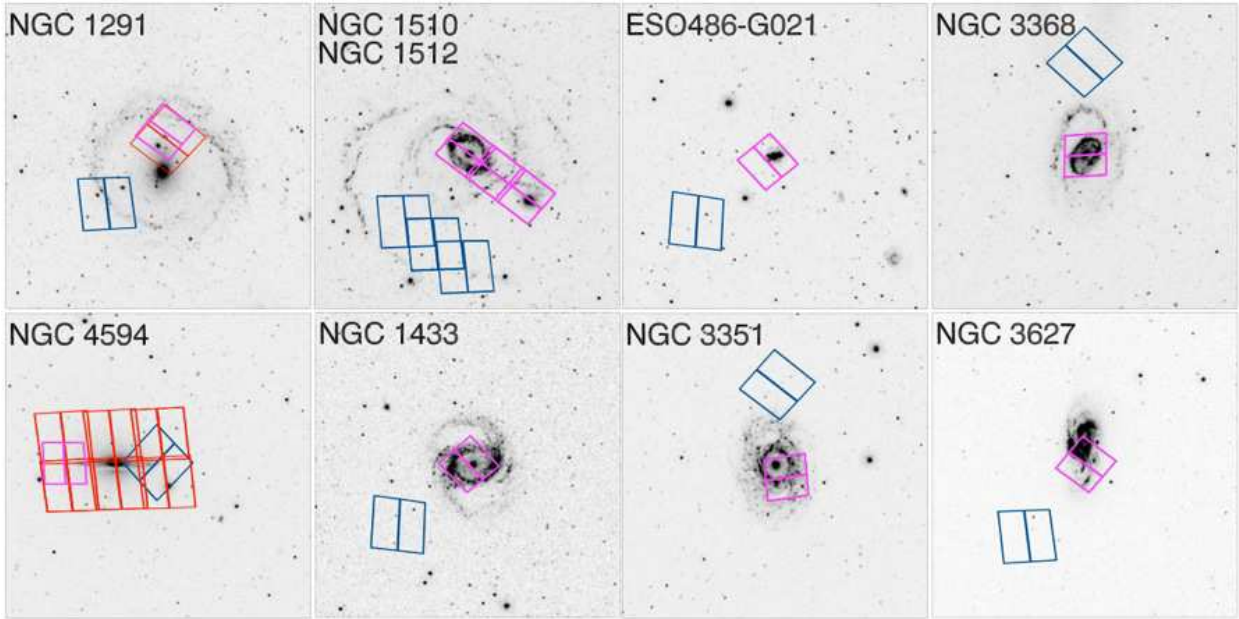}
\caption{Footprints of the WFC3 primary observations (magenta), ACS parallels (blue), and, when relevant, of the archival ACS images (red), for the 
50 LEGUS galaxies. The footprints are overlaid on the GALEX NUV images of the galaxies, with  20$^{\prime}\times$20$^{\prime}$ size, North up, East left.  
In the few cases of neighboring/interacting galaxies, two galaxies are shown on a panel (e.g., NGC1510/NGC1512; NGC5194/NGC5195). 
\label{fig3a}}
\end{figure}

\clearpage 
\begin{figure}
\figurenum{3b}
\plotone{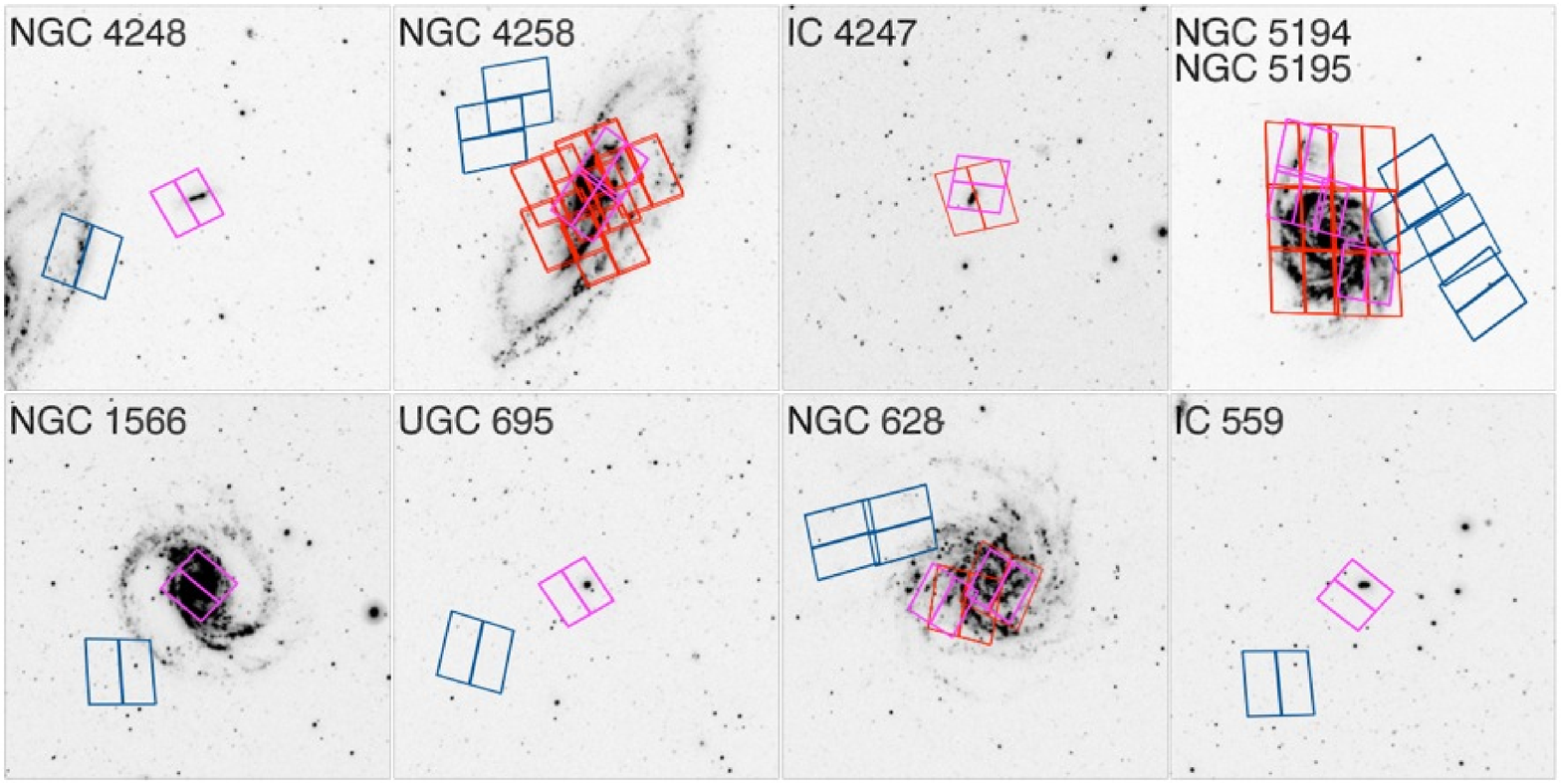}
\caption{Continued.
\label{fig3b}}
\end{figure}

\clearpage 
\begin{figure}
\figurenum{3c}
\plotone{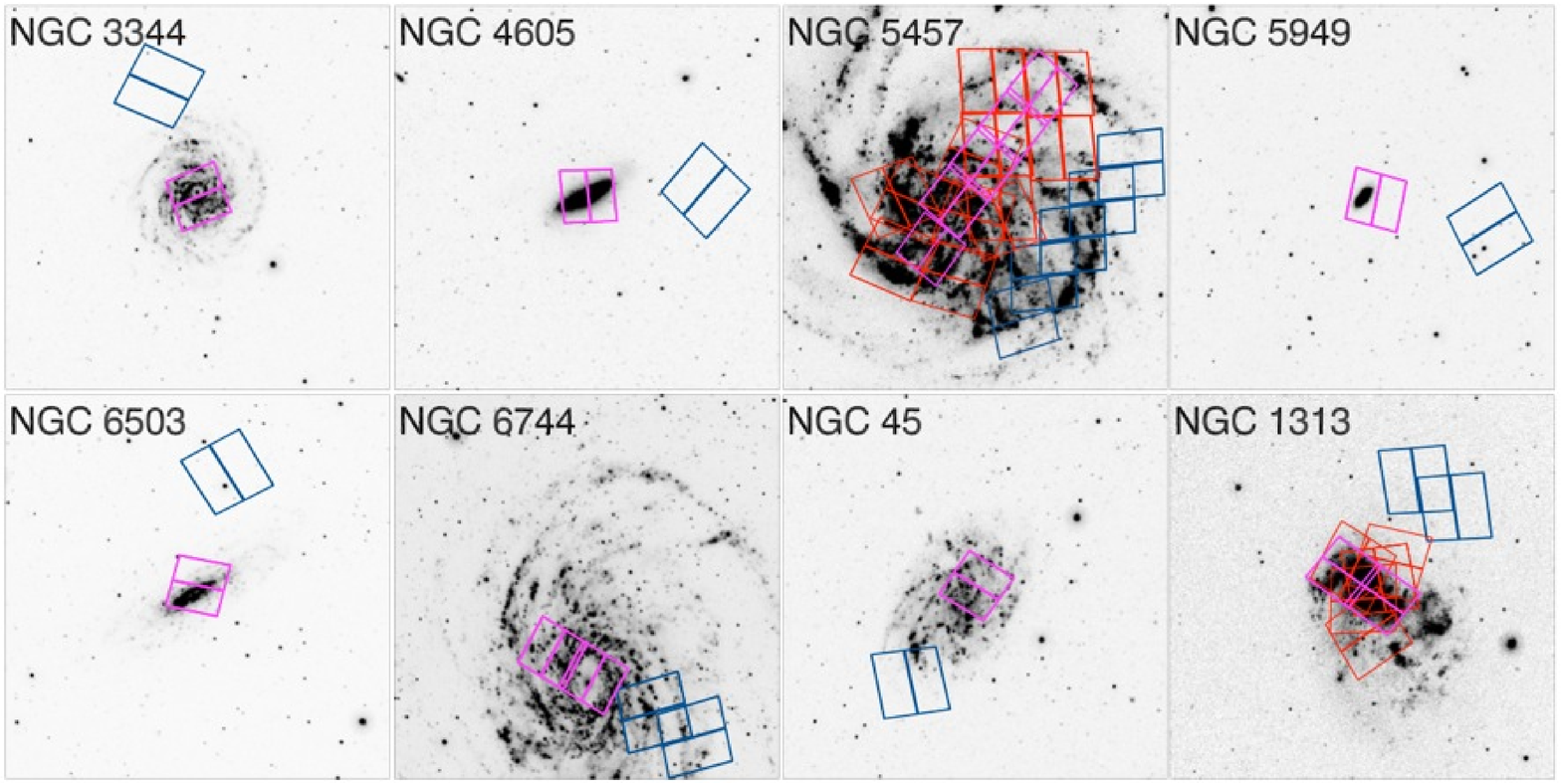}
\caption{Continued.
\label{fig3c}}
\end{figure}

\clearpage 
\begin{figure}
\figurenum{3d}
\plotone{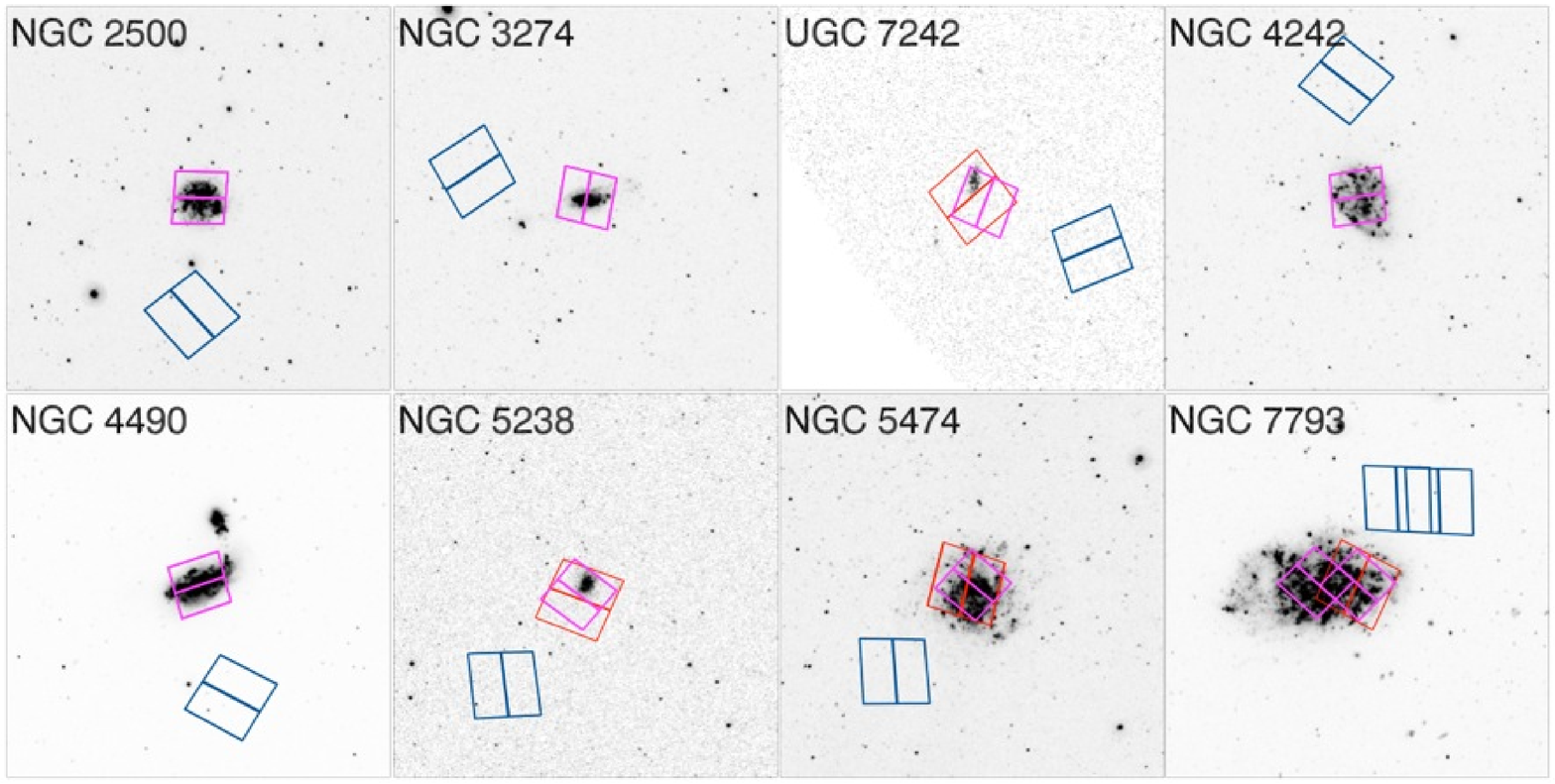}
\caption{Continued.
\label{fig3d}}
\end{figure}

\clearpage 
\begin{figure}
\figurenum{3e}
\plotone{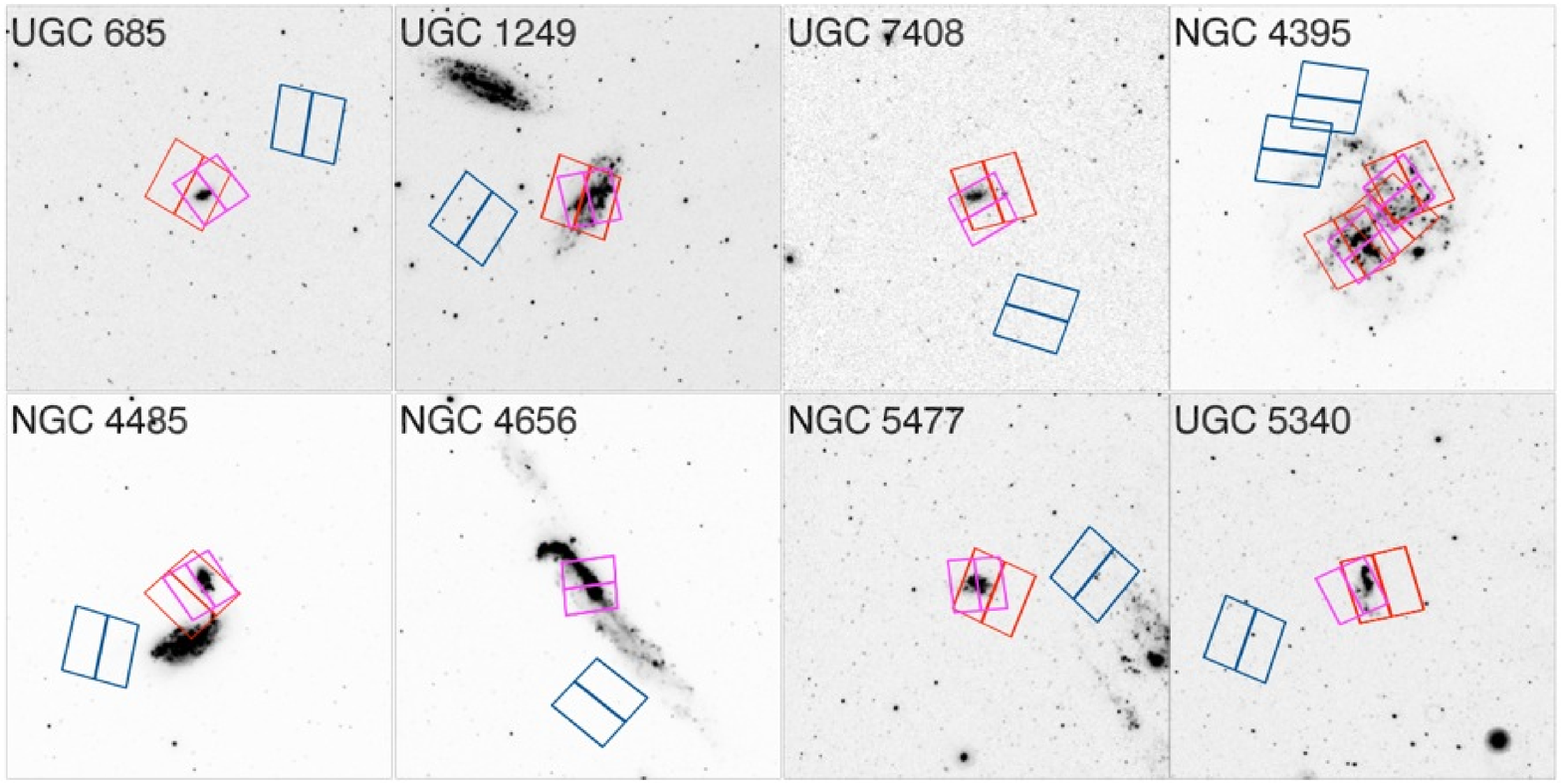}
\caption{Continued.
\label{fig3e}}
\end{figure}

\clearpage 
\begin{figure}
\figurenum{3f}
\plotone{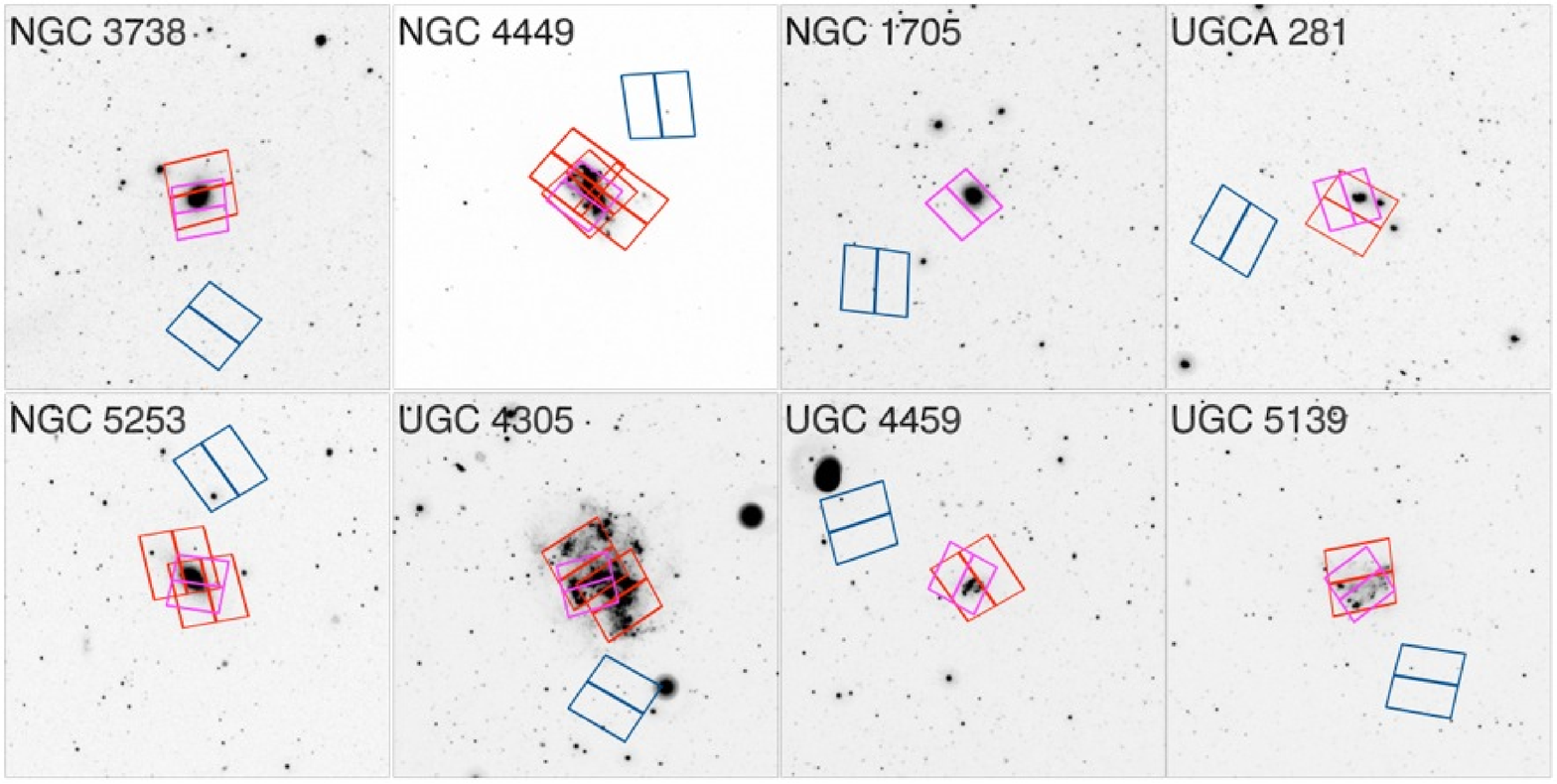}
\caption{Continued.
\label{fig3f}}
\end{figure}

\clearpage 
\begin{figure}
\figurenum{4}
\plotone{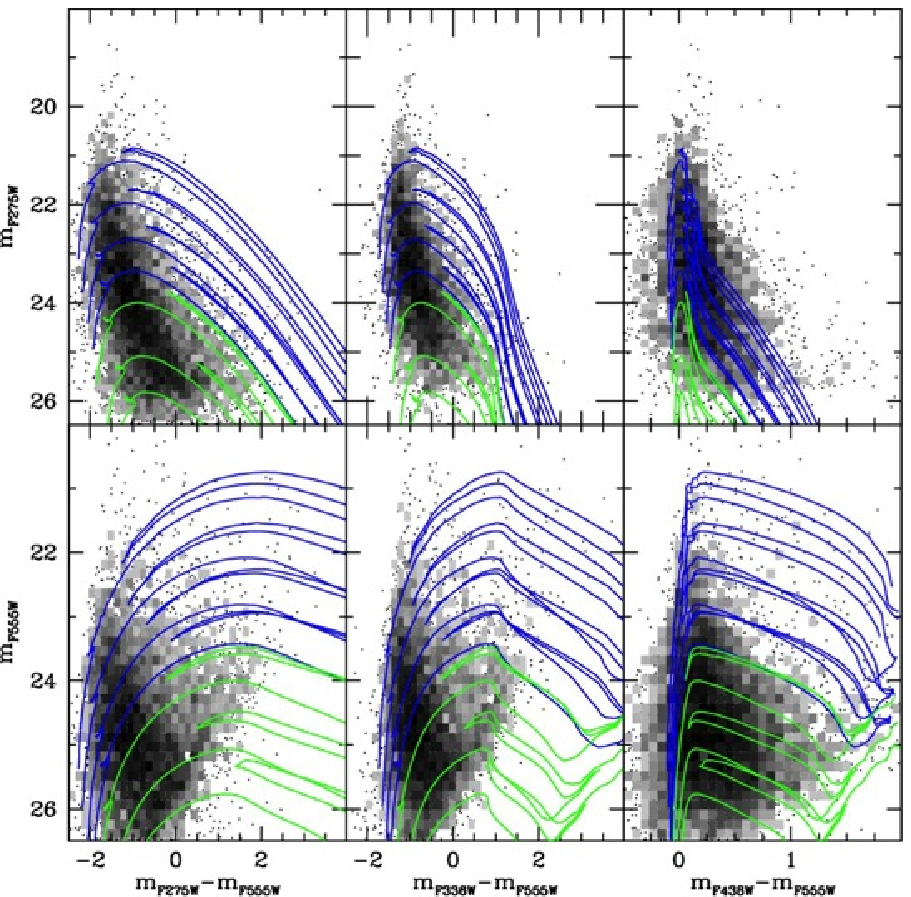}
\caption{Hess diagram of the galaxy NGC6503 showing the NUV (top) and V (bottom) on the vertical axis, as a function of X-V colors, 
where X=NUV,U,B. Photometry is in Vega magnitudes. The data are shown as  a black/gray density plot. The blue and green lines are 
Padova evolutionary tracks \citep{Girardi2010} of stars at a range of masses, for two distinct values of the metallicity; from the topmost track to the bottom one, 
masses and metallicities are:  20, 15, 12, and 10, M$_{\odot}$ with Z=0.017 (slightly--above solar metallicity) in blue and 
8, 7, and 6 M$_{\odot}$ with Z=0.008 in green. A reddening of E(B$-$V)=0.2 has been assumed for the tracks.  
\label{fig4}}
\end{figure}

\clearpage 
\begin{figure}
\figurenum{5}
\plotone{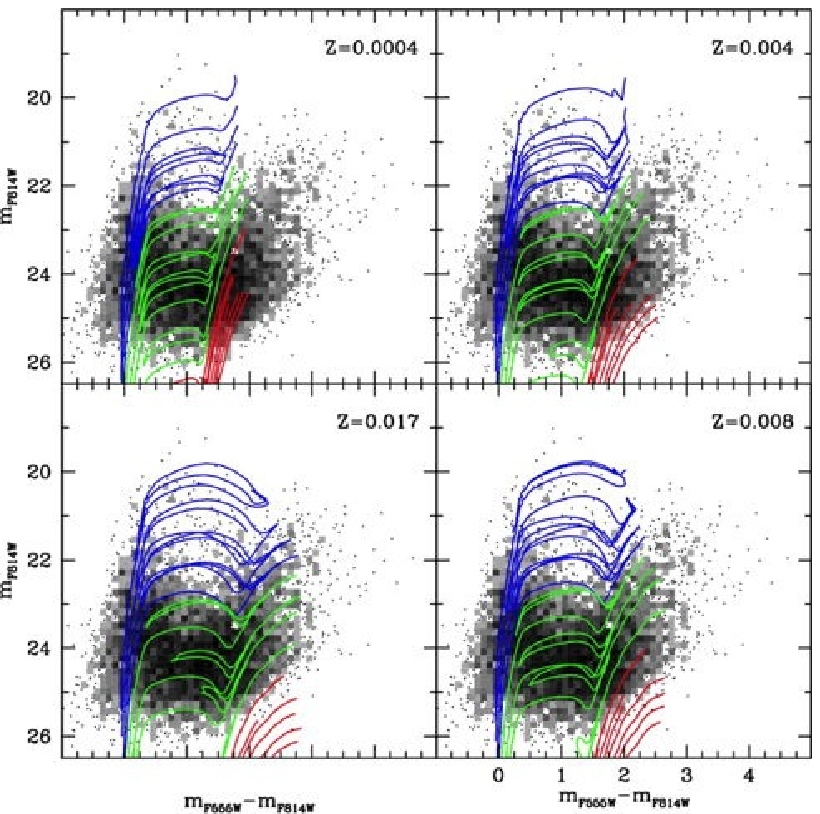}
\caption{The same as Figure~\ref{fig4}, but for the I band versus V--I. The same data are shown in each panel, but they are compared with 
models tracks at different metallicity values. Counter--clock--wise, from bottom--left to top--left: 
slightly--above--solar metallicity (Z=0.017); LMC--like metallicity (Z$_{LMC}\sim$0.008); 
SMC--like metallicity ((Z$_{SMC}\sim$0.004); and metallicity $\sim$1/35th solar. Tracks are in blue for stars with masses M$\ge$10~M$_{\odot}$, in green for 
2.5~M$_{\odot}<$M$\le$8~M$_{\odot}$, and red for M$\le$2.5~M$_{\odot}$.
\label{fig5}}
\end{figure}

\clearpage 
\begin{figure}
\figurenum{6}
\plotone{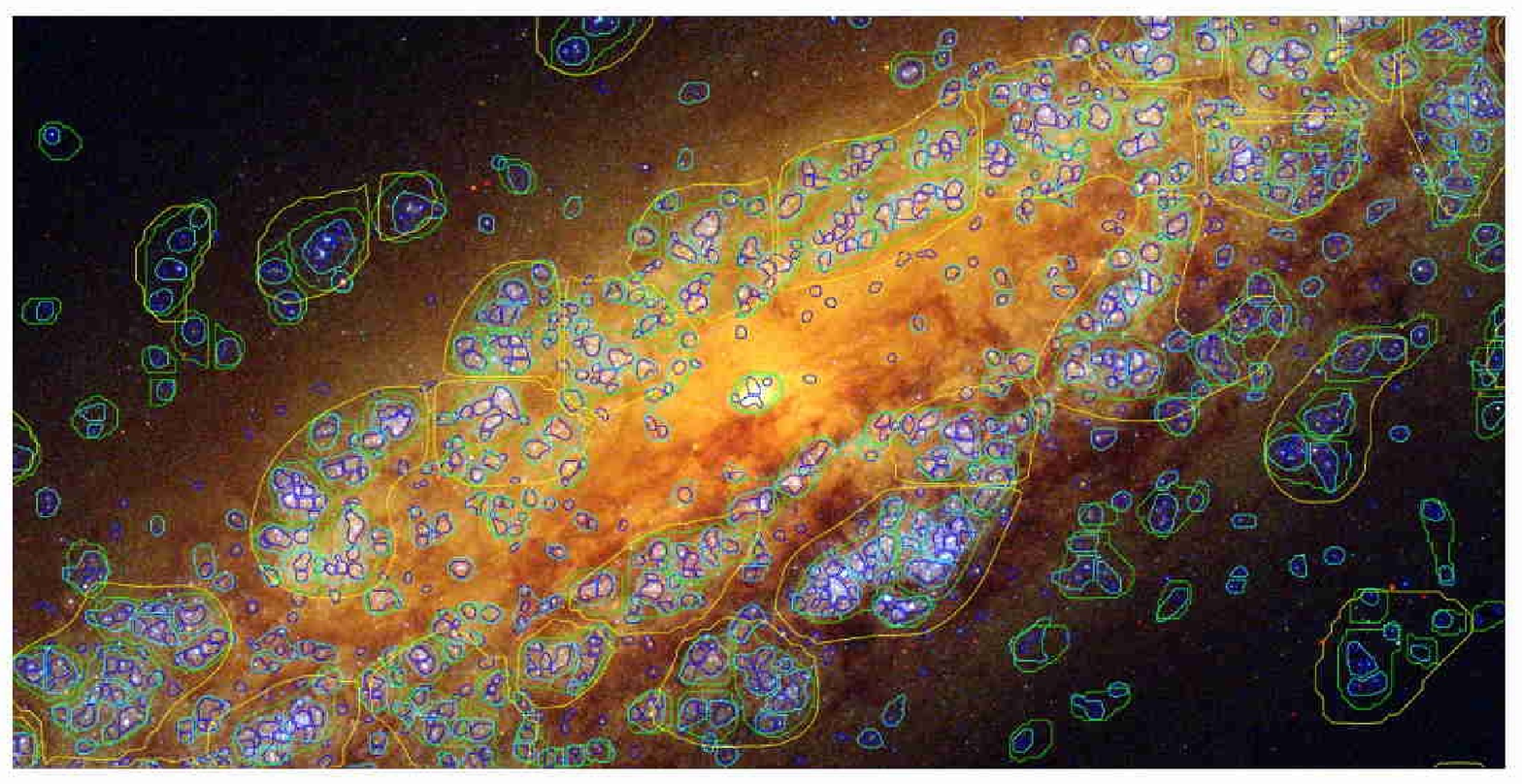}
\caption{
The hierarchical structures traced by the UV--bright stars in NGC6503 are shown on a three--color image (UV,B,I) of the galaxy. The UV--bright
population has been selected from the CMD of the galaxy, using the region delimited by $-$2$\le$NUV$-$U$\le$2~mag and brighter than absolute magnitude
M$_{NUV}-$2.5~mag. The four color contours (blue, cyan, green, and yellow, in order of increasing physical scale, separated by a factor 2 in smoothing kernel FWHM) delimit regions having significant difference in the
smoothed surface density of the UV--bright stars between one scale and the next. The largest regions have sizes $\sim$700~pc. Our method links together any spatially associated over-densities detected at arbitrary scale into composite hierarchical structures.  The field of view of the
image is $\sim$3.3~kpc$\times$1.6~kpc.
\label{fig6}}
\end{figure}

\clearpage 
\begin{figure}
\figurenum{7}
\plotone{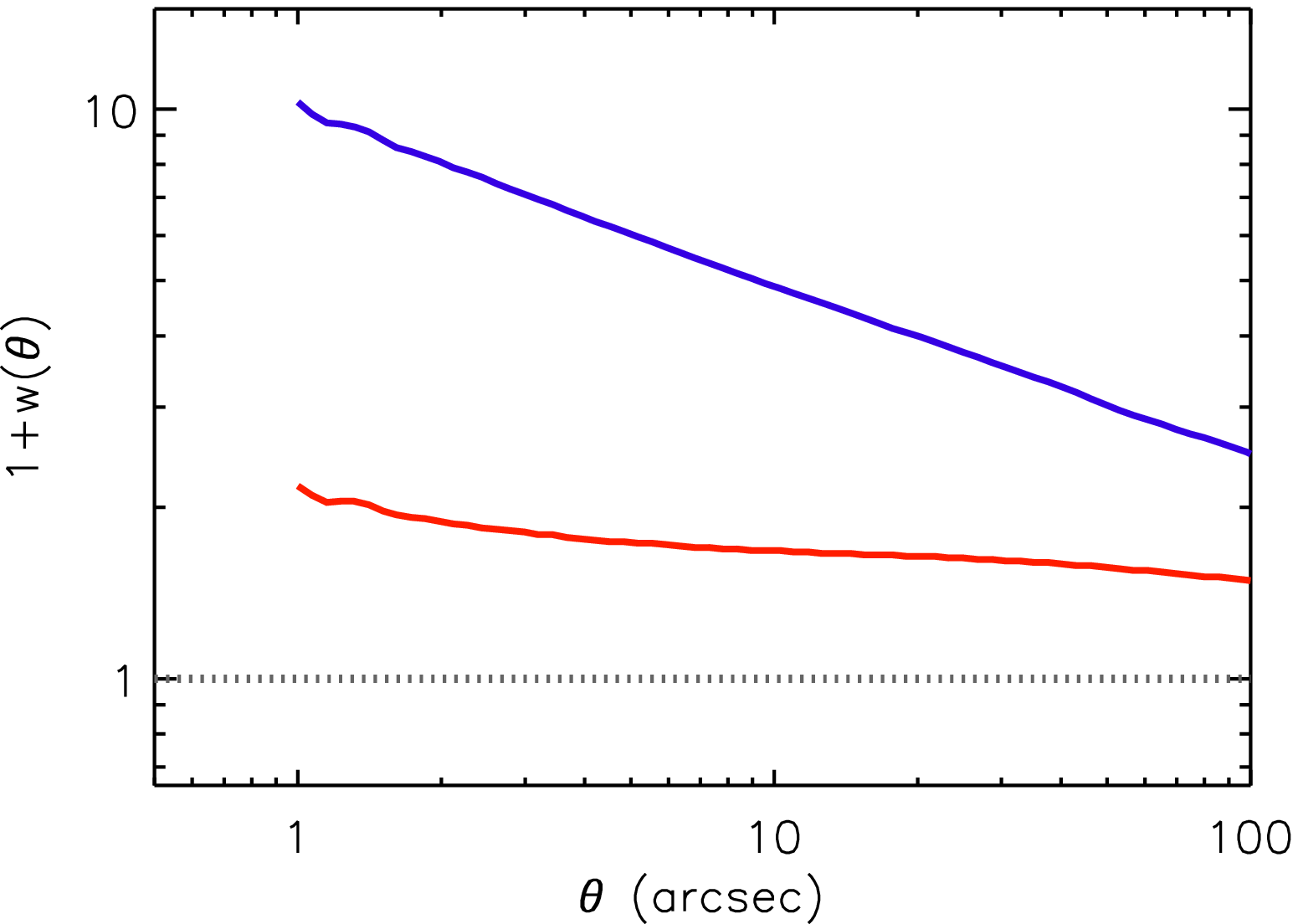}
\caption{The two--point angular correlation function of the young stars ($< 100$\,Myr; blue line) and old stars
($> 500$\,Myr; red line) for the whole extent of  NGC\,6503. The horizontal grey dotted line is the expected two--point angular correlation function 
of a randomly distributed population. The monotonically decreasing functions imply that the stellar distributions follow a hierarchical pattern, with the 
young stars more strongly clustered than the old stars across galactic scales up to projected sizes of at least 100\arcsec\ (equivalent to $\sim$\,2.75\,kpc). .
\label{fig7}}
\end{figure}

\clearpage 
\begin{figure}
\figurenum{8}
\plottwo{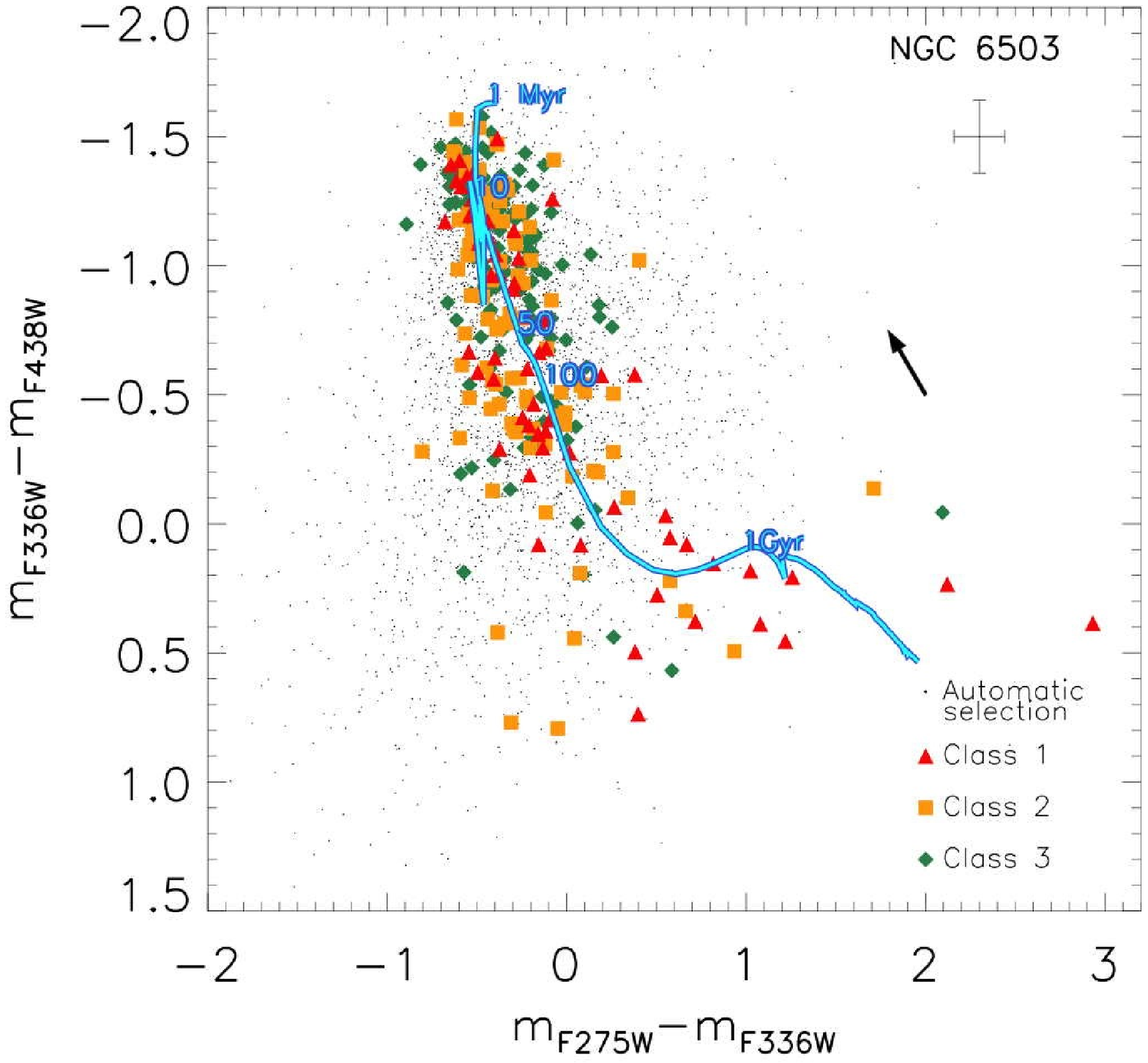}{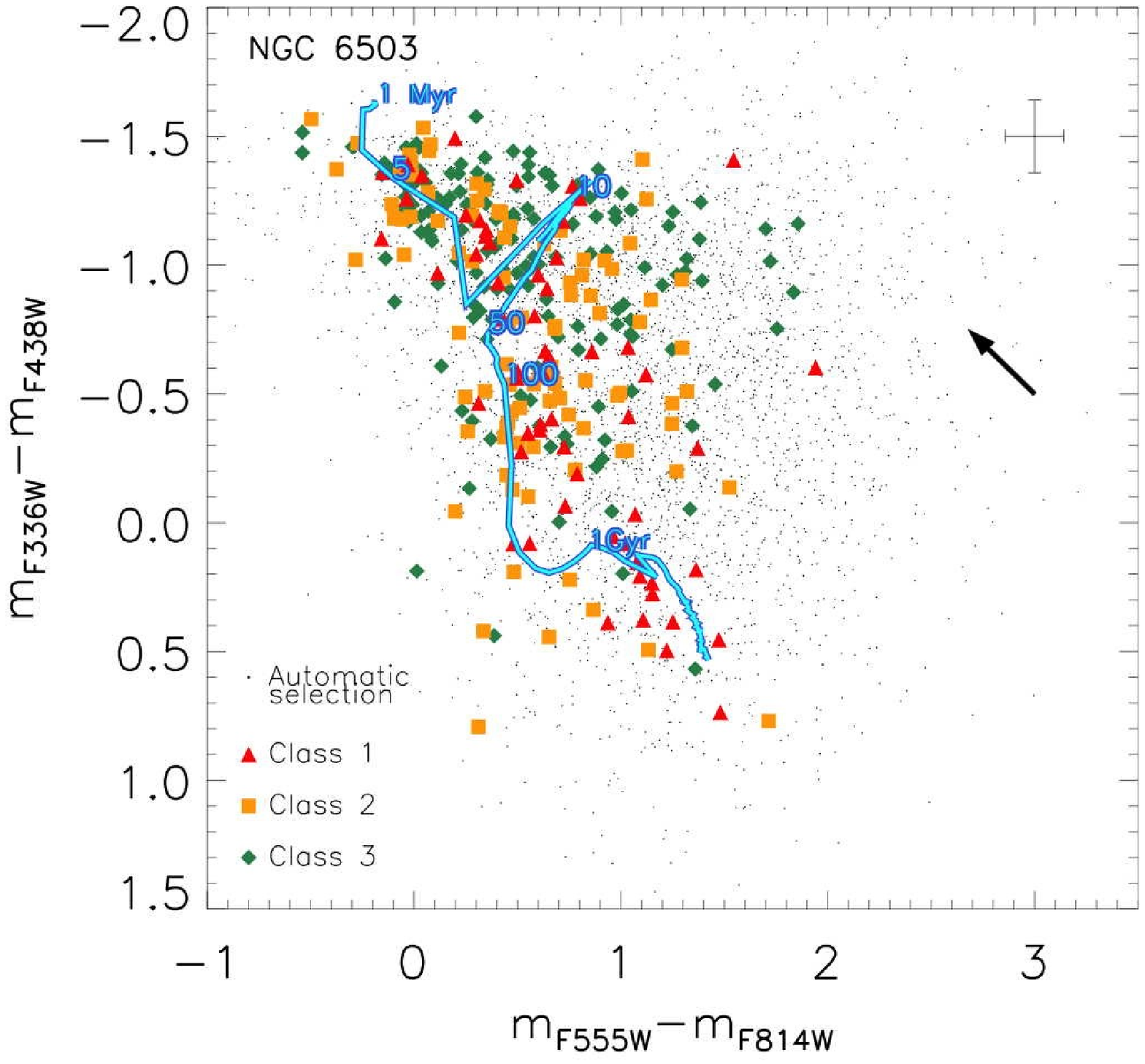}
\caption{The color--color diagrams (CCDs) of the star cluster candidates (small grey dots, about 4600 in total) and the confirmed clusters (colored dots, about 290 in total) in NGC6503. The CCD in the left--hand--side panel includes the NUV--U color along the x--axis, while the CCD in the right--hand--side panel uses the classical U--B versus V--I  axes. The color dots are coded according to the class assigned to the cluster, 1, 2, or 3 (see text), and represent our `high fidelity' star cluster sample. Models of evolving single--age stellar populations are also reported for comparison, with a light--blue curve, and with a few ages indicated between 1~Myr and 1~Gyr. The average value of the error bar is shown in each panel. The black arrow is the extinction vector, and shows the direction in which the colors would change if corrected for dust attenuation \citep[assuming the attenuation curve of][]{Calzetti2000}. The 
length of the arrow corresponds to a color excess E(B--V)=0.2, corresponding to A$_V\sim$0.8~mag.
\label{fig8}}
\end{figure}

\clearpage 
\begin{figure}
\figurenum{9}
\plottwo{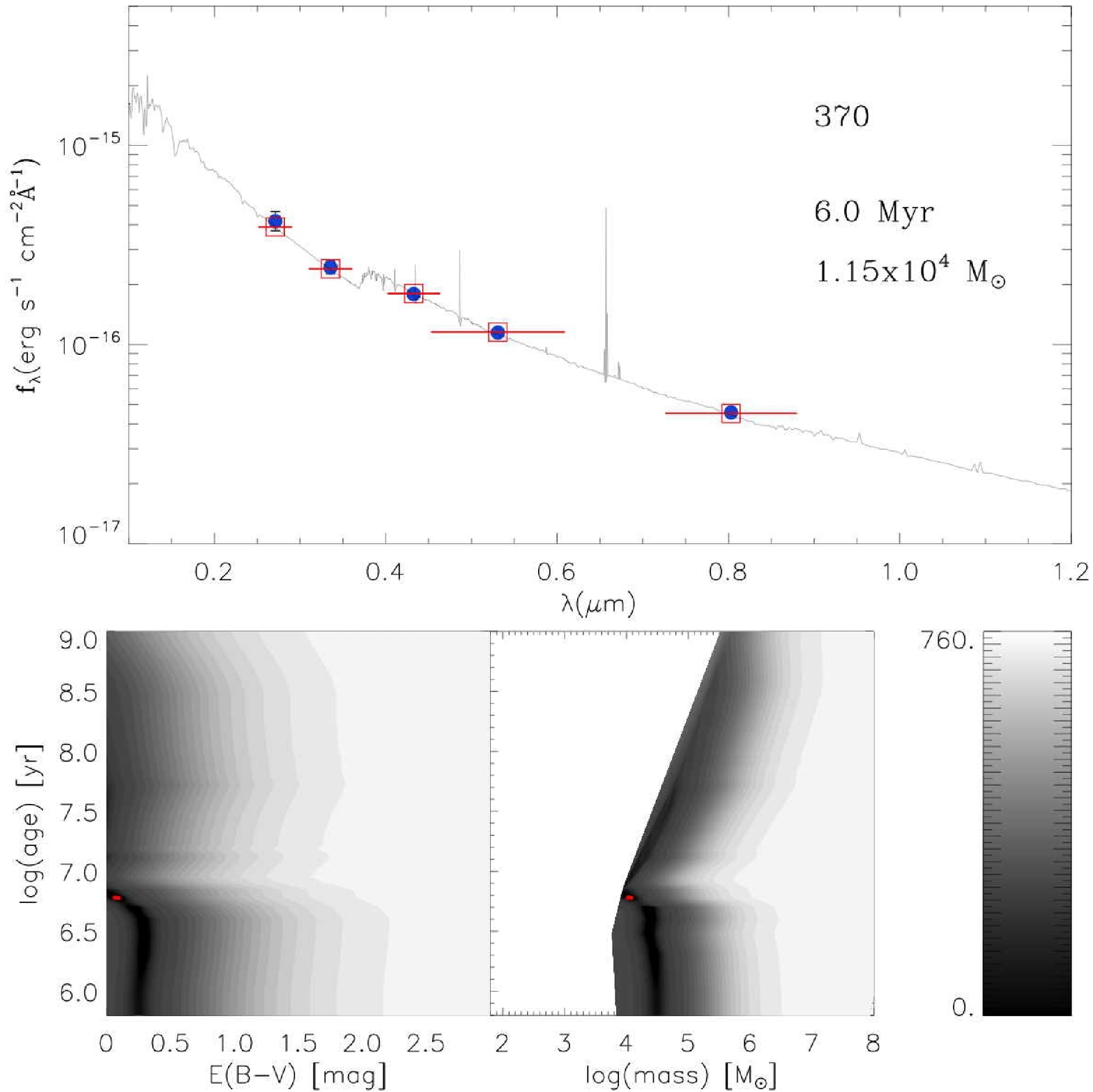}{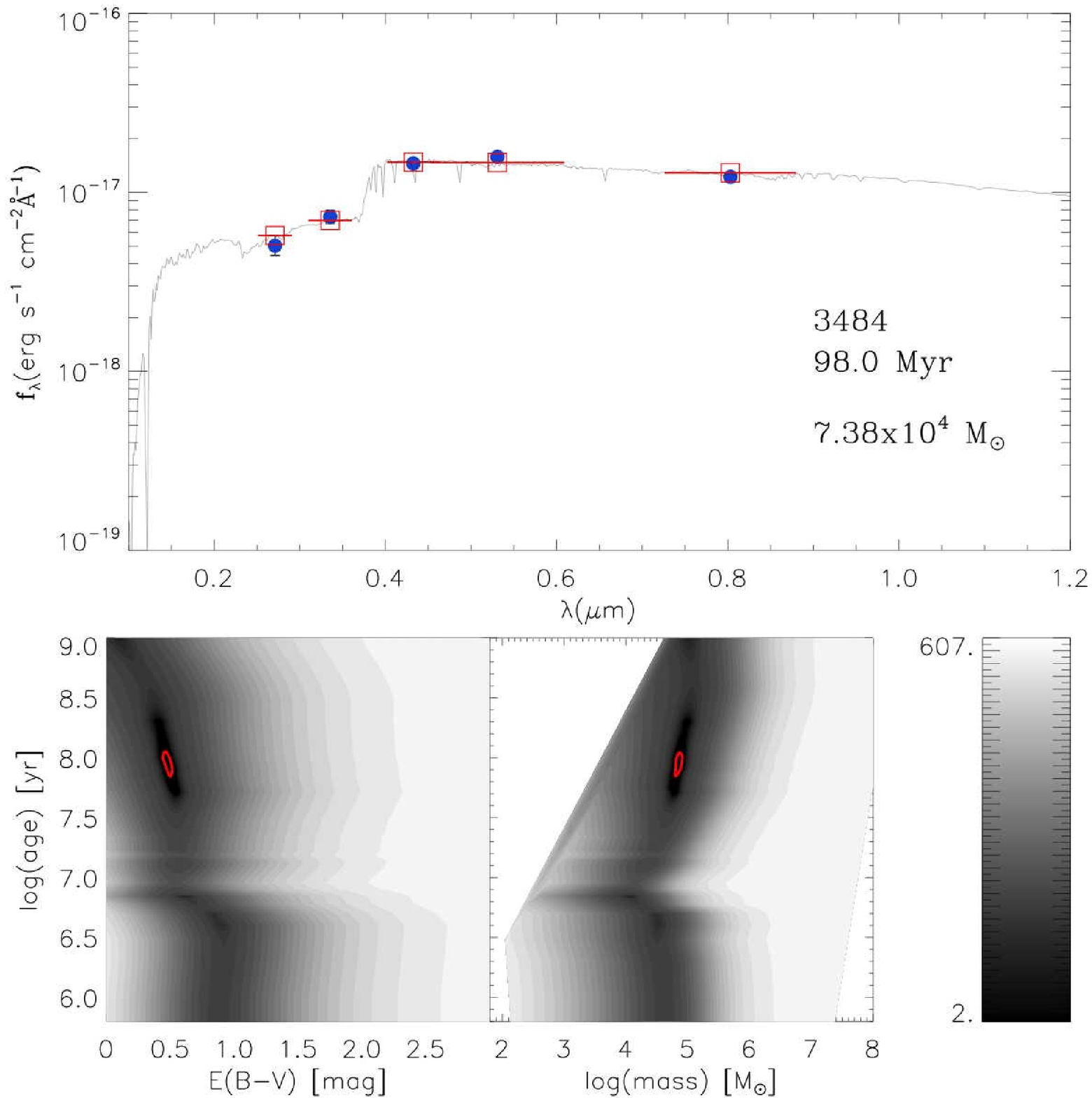}
\caption{SED fits of two class~1 star clusters, \# 370 (left panels) and \# 3484 (right panels). The best fits to the 5 LEGUS photometric bands were performed 
with the algorithm described in \citet{Adamo2010} and the error analysis described in \citet{Adamo2012}, that implement the Yggdrasil models with a Kroupa 
IMF \citep{Kroupa2001}, solar metallicity, and Padova isochrones. Nebular continuum and lines are included with a covering factor of 0.5 in these fits \citep{Zackrisson2011}. Cluster 
\# 370 has a best fit age around 6~Myr and \# 3484 around 100~Myr, and both have masses $>$10$^4$~M$_{\odot}$. For each cluster, the top panel shows the 
observed SED (red squares with error bars) and the best-fit synthetic spectrum$+$photometry (continuous line and blue triangles). The two panels below the 
SED panels show the $\chi^2$ distribution in age, mass, and color excess E(B--V), with the scale given by the grey scale to the right of each set of panels. 
The 68\% confidence level regions around the minimum $\chi^2$ values are shown as red contours in the age--versus--mass and age--versus--E(B--V) distributions. 
\label{fig9}}
\end{figure}

\end{document}